\documentclass[twoside,twocolumn,tightenlines,superscriptaddress,floatfix,showpacs,aps,prb]{revtex4}
\usepackage{graphicx}
\usepackage{wasysym}
\usepackage{amssymb}
\usepackage{bbold}
\usepackage[makeroom]{cancel}
\newcommand{\be}{\begin{equation}}
\newcommand{\ee}{\end{equation}}
\usepackage[unicode=true,pdfusetitle,
 bookmarks=true,bookmarksnumbered=false,bookmarksopen=false,
 breaklinks=false,pdfborder={0 0 1},backref=false,colorlinks=false]
 {hyperref}
\hypersetup{
 colorlinks,linkcolor=blue,citecolor=blue,urlcolor=blue}

\begin{document}

\preprint{APS/123-QED}

\title{Orbital Hall effect in mesoscopic devices}

\author{Diego B. Fonseca}

\author{Lucas L. A. Pereira}

\author{Anderson L. R. Barbosa}
\email{anderson.barbosa@ufrpe.br}
\affiliation{Departamento de F\'{\i}sica, Universidade Federal Rural de Pernambuco, 52171-900, Recife, PE, Brazil}

\date{\today}

\begin{abstract}
We investigate the orbital Hall effect through a mesoscopic device with momentum-space orbital texture that is connected to four semi-infinite terminals embedded in the Landauer-B\"uttiker configuration for quantum transport. {We present analytical and numerical evidence that the orbital Hall current exhibits mesoscopic fluctuations, which can be interpreted in the framework of random matrix theory (RMT) (as with spin Hall current fluctuations). The mesoscopic fluctuations of orbital Hall current display two different amplitudes of 0.36 and 0.18 for weak and strong spin-orbit coupling, respectively. The amplitudes are obtained by analytical calculation via RMT and are supported by numerical calculations based on the tight-binding model. Furthermore, the orbital Hall current fluctuations lead to two relationships between the orbital Hall angle and conductivity. Finally, we confront the two relations with experimental data of the orbital Hall angle, which shows good concordance between theory and experiment.}
\end{abstract}


\maketitle


\section{Introduction} 

The Spin Hall effect (SHE) is one of the most prominent phenomena observed in spintronics, which allows us to convert a longitudinal charge current to a transversal spin Hall current (SHC) \cite{pereldois,spinhallh,Kato1910,PhysRevLett.94.047204,PhysRevB.72.075361,PhysRevB.74.035340,GORINI2022,RevModPhys.87.1213,coloquiumspintronics}. 
Spin-orbit coupling (SOC) is the key behind the SHE because it lets us control spin transport properties without magnetic materials. 
Furthermore, the spin Hall angle (SHA) is an important parameter that is commonly used to quantify a material's ability to convert charge-to-spin
currents. SHA is defined as the ratio between the SHC and the charge current, and has been measured in various heavy metals---that is, metals with strong SOC, such as Pt \cite{PhysRevB.94.060412}, and W \cite{doi:10.1063/1.4753947}--- and in two-dimensional materials---such as graphene \cite{10.1038/nphys2576,10.1038/ncomms5748}.

Much attention has been given to the orbital Hall effect (OHE), which is a phenomenon of orbitronics \cite{2023Natur.619...38R,PhysRevLett.95.066601,PhysRevB.77.165117,PhysRevLett.123.236403,PhysRevLett.121.086602,PhysRevB.98.214405,PhysRevMaterials.5.074407,PhysRevB.101.161409,PhysRevB.101.075429,PhysRevB.102.035409,PhysRevLett.126.056601,PhysRevLett.130.116204,PhysRevB.103.085113,PhysRevB.103.195309,PhysRevResearch.2.013177,PhysRevLett.128.067201,PhysRevB.105.104434,PhysRevLett.125.177201,salvadorsanchez2022generation,PhysRevLett.128.176601,Lee2021,https://doi.org/10.48550/arxiv.2109.14847,https://doi.org/10.48550/arxiv.2202.13896,PhysRevResearch.4.033037,https://doi.org/10.48550/arxiv.2204.01825,PhysRevB.104.245204,PhysRevB.107.134423,Go_2021,KIM2022169974}.
As shown by D. Go $et.al.$ [\onlinecite{PhysRevLett.121.086602}], we can convert a longitudinal charge current to a transversal orbital Hall current (OHC) in centrosymmetric systems with momentum-space
orbital texture, even when the orbital angular momentum is quenched in equilibrium. 
A remarkable feature of OHE is that it is independent of SOC, in contrast with SHE.
Therefore, we can consider the OHE to be more fundamental than the SHE \cite{PhysRevLett.121.086602,PhysRevB.103.085113}.
Similar to SHA, the orbital Hall angle (OHA) quantifies a material's ability to convert charge-to-orbital currents and was measured in light metals as Ti \cite{https://doi.org/10.48550/arxiv.2109.14847,https://doi.org/10.48550/arxiv.2202.13896} and Cr \cite{PhysRevResearch.4.033037} (i.e., metals with weak SOC) and heavy metals as W \cite{https://doi.org/10.48550/arxiv.2202.13896} and Pt \cite{PhysRevResearch.4.033037}.

\begin{figure}
\includegraphics[scale = 0.3]{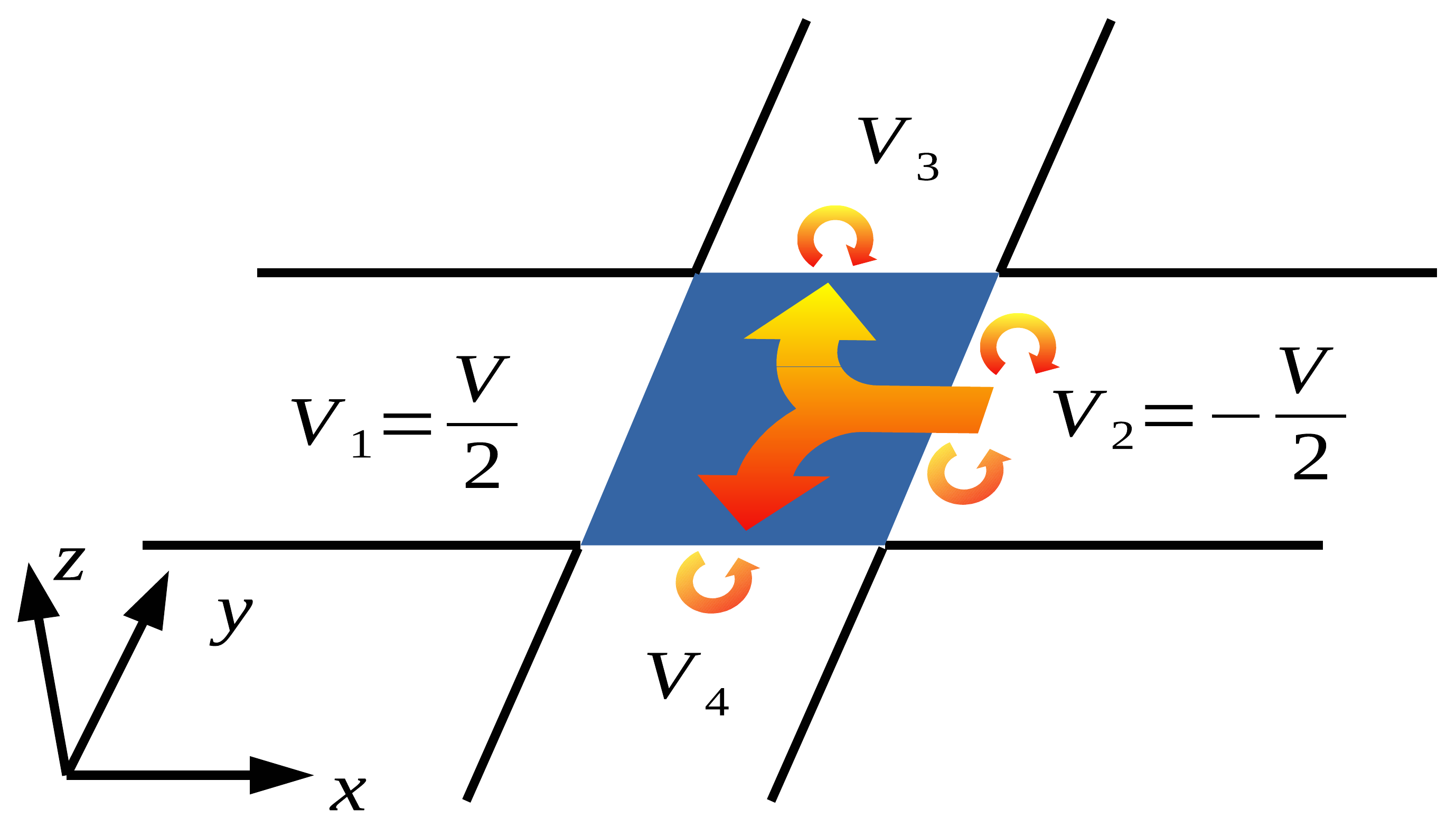}
\caption{OHE through a disordered mesoscopic device with space-moment orbital texture (blue) connected to four semi-infinite terminals subjected to voltages $V_i$. The SOC may or may not be included in the mesoscopic device.}\label{sample}
\end{figure}

{As shown in the 1980s \cite{doi:10.1080/00018738600101921}, the charge current through the mesoscopic diffusive device in the linear regime at low temperature exhibits mesoscopic fluctuations, which are theoretically interpreted within the framework of random matrix theory (RMT) \cite{RevModPhys.69.731}.
Therefore, in the early SHE experiments \cite{Kato1910,PhysRevLett.94.047204}, the interest in whether the SHC exhibits mesoscopic fluctuations appeared.
The mesoscopic fluctuations of SHC were numerically demonstrated by Ref. [\onlinecite{PhysRevLett.97.066603}] and confirmed analytically via RMT by Ref. [\onlinecite{PhysRevLett.98.196601}].
However, the SHC fluctuations (SHCF) have never been confirmed experimentally because the SHC is only measured indirectly via the inverse spin Hall effect \cite{doi:10.1063/1.2199473,doi:10.1063/1.1855251,PhysRevLett.115.226601,doi:10.1063/1.5010973}.
The connection between SHCF and SHE experiments was made by Refs. [\onlinecite{PhysRevB.102.041107,doi:10.1063/5.0107212}], who show that the SHCF lead to a relationship between the maximum SHA deviation $\Theta_{\text{SH}}$ and dimensionless longitudinal conductivity $\sigma =N l_e/L$, where $N$, $L$ and $l_e$ are the number of propagating wave modes, longitudinal device length, and free electron path, respectively, which is given by
$\Theta_{\text{SH}} \times \sigma = 0.18$. Therefore, the question that arises in the early OHE experiments is \cite{https://doi.org/10.48550/arxiv.2109.14847,https://doi.org/10.48550/arxiv.2202.13896,Lee2021,PhysRevResearch.4.033037}: does the OHC exhibit mesoscopic fluctuations?}

In this work, we study the OHE through a mesoscopic device with momentum-space orbital texture that is connected to four semi-infinite terminals that are embedded in the Landauer-B\"uttiker configuration for quantum transport, as shown in Fig.(\ref{sample}). 
{Using analytical calculations via RMT and numerical calculations based on the tight-binding model for a square lattice with four orbitals, we report mesoscopic fluctuations of OHC with different amplitudes for light and heavy metals. Our finds are valid for ballistic chaotic and mesoscopic diffusive devices in the limit when the mean dwell time of the electrons is much longer than the time needed for ergodic exploration of the phase space, $\tau_{dwell} \gg \tau_{erg}$. Furthermore, the OHC fluctuations (OHCF) lead to two relationships between the maximum OHA deviation and dimensionless longitudinal conductivity. 
Finally, we confront the two relations with experiment data of Refs. [\onlinecite{https://doi.org/10.48550/arxiv.2109.14847,https://doi.org/10.48550/arxiv.2202.13896,PhysRevResearch.4.033037}], and conclude the compatibility between theory and experiments.}

\section{Orbital Hall effect} 

We designed the OHE setup through a mesoscopic device with orbital angular momentum and spin degrees of freedom that is connected to four semi-infinite terminals that are submitted to voltages $V_i$, Fig.(\ref{sample}).
From the Landauer-B\"uttiker model, the OHC (SHC) through the $i$th terminal in the linear regime at low temperature is
\begin{equation}
    I^{o(s)}_{i,\eta} = \frac{e^2}{h}\sum_{j} \tau_{ij,\eta}^{o(s)} \left( V_i - V_j \right),\label{IOS}
\end{equation} 
where the orbital (spin) transmission coefficient is calculated from the transmission and reflection blocks of scattering matrix $\mathcal{S}=\left[\mathcal{S}_{ij}\right]_{i,j=1,\dots,4}$
$$
\tau_{ij,\eta}^{o(s)} =\textbf{Tr}\left[\left(\mathcal{S}_{ij}\right)^{\dagger} \mathcal{P}^{o(s)}_\eta \mathcal{S}_{ij}\right].
$$
The matrix $\mathcal{P}^{o(s)}_\eta = \mathbb{1}_N \otimes l^\eta \otimes \sigma^0$ $\left(\mathbb{1}_N \otimes l^0 \otimes \sigma^\eta\right)$ is a projector, where $\mathbb{1}_N$ is a identity matrix with dimension $N\times N$. The dimensionless integer $N$ is the number of propagating wave modes in the terminals, proportional to the terminal width ($W$) and the Fermi vector ($k_F$) through the equation $N = k_F W/\pi$.  The index $\eta = \{0,x,y,z\}$, $l^0=(l^\eta)^2$, $\sigma^0=(\sigma^\eta)^2$, and $l^\eta$ and $\sigma^\eta$ are orbital angular momentum and Pauli matrices, respectively. Therefore, the charge current is defined by $\eta=0$, while OHC (SHC) by $\eta=\{x,y,z\}$.

The pure OHC (SHC) $$I^{o(s)}_{i,z}=I^{\circlearrowleft(\uparrow)}_{i}-I^{\circlearrowright(\downarrow)}_{i}, \quad i=3,4$$ can be obtained by assuming that the charge current vanishes in the transverse terminals, $$I^{c}_{i,0}=I^{\circlearrowleft(\uparrow)}_i+I^{\circlearrowright(\downarrow)}_i=0, \quad i=3,4,$$ while the charge current is conserved in the longitudinal terminals\cite{PhysRevB.72.075361,Nikolic_2007,PhysRevLett.98.196601} $$I^c_{1,0}=-I^c_{2,0}=I^c.$$ By applying these conditions to Eq.(\ref{IOS}), we obtain
\begin{eqnarray}
I^{o(s)}_{i,\eta} = \frac{e^2}{h}\left[\left(\tau_{i2,\eta}^{o(s)}-\tau_{i1,\eta}^{o(s)}\right)\frac{V}{2}
- \tau_{i3,\eta}^{o(s)}V_3 + \tau_{i4,\eta}^{o(s)}V_4\right],
\label{Is}
\end{eqnarray}
for $i=3,4$, where $V$ is a constant potential difference between longitudinal terminals, and $V_{3,4}$ is the transversal terminal voltage. The nature of the OHC in Eq. (\ref{Is})  is a charge current moving through the orbital degrees of freedom projected by $\mathcal{P}^{o}_\eta$. A detailed demonstration of Eq. (\ref{Is}) can be found in Appendix \ref{A}.

We consider a mesoscopic device Fig.(\ref{sample}), which allows us to analyse the OHE in the framework of RMT \cite{RevModPhys.69.731}. Without an external magnetic field applied, the mesoscopic device preserves time-reversal symmetry. Therefore, the scattering matrix is described by the circular orthogonal ensemble (COE) when SOC is absent (light metals) and the circular symplectic ensemble (CSE) when SOC is strong (heavy metals). Consequently, we can calculate the average and variance of the OHC (\ref{Is}) by applying the method of Ref. [\onlinecite{doi:10.1063/1.531667}]. The calculation is valid for the ballistic chaotic and mesoscopic diffusive devices in the limit when the mean dwell time of the electrons is much longer than the time needed for ergodic exploration of the phase space, $\tau_{dwell} \gg \tau_{erg}$ \cite{RevModPhys.69.731,PhysRevLett.98.196601}.

Without loss of generality, we consider a mesoscopic device with four orbitals (i.e., $s$ and $p$ orbitals) and $\eta = z$. In this case
$$
l^z=\left[\begin{array}{cccc}
0&0&0 &0\\
0&0&-i&0\\
0&i&0&0\\
0&0&0&0
\end{array}\right], \; \sigma^z=\left[\begin{array}{cc}
1&0\\
0&-1
\end{array}\right],
$$
and the scattering matrix has dimension $32N \times 32N$.
To perform the average of Eq.(\ref{Is}), we must take the experimental regime of interest; that is, when the sample has a large thickness $N\gg1$. Therefore, we can assume the central limit theorem (CLT) \cite{Reichl:101976} and expand Eq.(\ref{Is}) in the function of $N$ \cite{RevModPhys.69.731}. By applying the method of Ref. [\onlinecite{doi:10.1063/1.531667}] in Eq.(\ref{Is}), we find
\begin{eqnarray}
\left\langle I^{o(s)}_{\eta}\right\rangle=0,\label{mean}
\end{eqnarray}
for COE and CSE. The SHC average was previously calculated by [\onlinecite{PhysRevLett.98.196601,PhysRevB.86.235112,PhysRevB.93.115120}]. Eq.(\ref{mean}) implies a zero-mean Gaussian distribution, meaning all relevant information can be contained in OHC fluctuations. Therefore, we are interested in the OHC deviation because although the mean of one is zero, its mesoscopic fluctuations can be significant. 

In the usual way, the OHC variance is defined as $$
\textbf{var}[I^{o}_{i,\eta}]=\left\langle {I^{o}_{i,\eta}}^2\right\rangle-\left\langle I^{o}_{i,\eta}\right\rangle^2 = \left\langle  {I^{o}_{i,\eta}}^2\right\rangle,
$$ and by applying the method of Ref. [\onlinecite{doi:10.1063/1.531667}], we obtain
\begin{eqnarray}
\textbf{var}[I^{o}_{\eta}]=\left(\frac{e^2V}{h}\right)^2 \times \Bigg\{\begin{array}{cc}
\frac{2N\left(4N+1\right)}{\left(8N+1\right)\left(8N+3\right)} & \text{for COE}\\
\frac{N\left(8N-1\right)}{\left(16N-1\right)\left(16N-3\right)} & \text{for CSE}
\end{array},\label{IOSS}
\end{eqnarray}
for  $i=3,4$ and $\eta = \{x,y,z\}$.
When the sample has a large thickness  $N\gg1$, Eq. (\ref{IOSS}) goes to
\begin{eqnarray}
\textbf{var}[I^{o}_{\eta}]=\left(\frac{e^2V}{h}\right)^2 \times \left\{\begin{array}{cc}
\frac{1}{8} & \text{for COE}\\
\frac{1}{32} & \text{for CSE}
\end{array}\right.,\label{IOO}
\end{eqnarray}
for $i=3,4$ and $\eta = \{x,y,z\}$.
The OHC deviation is obtained from the OHC variance as
$$
\textbf{rms}[I^{o}_{\eta}]=\sqrt{\textbf{var}[I^{o}_{\eta}]}.\nonumber
$$
Then, we obtain
\begin{eqnarray}
\textbf{rms}[I^{o}_{\eta}]=\frac{e^2V}{h} \times \left\{\begin{array}{ccc}
\sqrt{\frac{1}{8}} & \approx 0.36 & \text{for COE}\\
\sqrt{\frac{1}{32}} & \approx0.18 & \text{for CSE}
\end{array},\right.\label{IO}
\end{eqnarray}
for $\eta = \{x,y,z\}$. Meanwhile, $\textbf{rms}[I^{s}_{\eta}]=0$ and $0.18$ for COE and CSE, respectively \cite{PhysRevLett.98.196601,PhysRevB.86.235112,PhysRevB.93.115120}. {Equation (\ref{IO}) is the first outcome of this work, which indicates that the OHC exhibits mesoscopic fluctuations, as with SHCF \cite{PhysRevLett.97.066603}. OHCF of light metals (COE) are consistent with the interpretation that the OHE is more fundamental than the SHE because it is independent of SOC. Furthermore, when the SOC is increased, the OHCF of light metals (COE) is decreased by a factor of 2 to the OHCF of heavy metals (CSE) because SOC breaks the spin-rotation symmetry. In this case, OHCF and SHCF exhibit the same mesoscopic fluctuations amplitude.}  A detailed demonstration of Eqs. (\ref{mean}) and (\ref{IOSS}) can be found in Appendix \ref{B}.

\begin{figure}
\includegraphics[scale = 0.7]{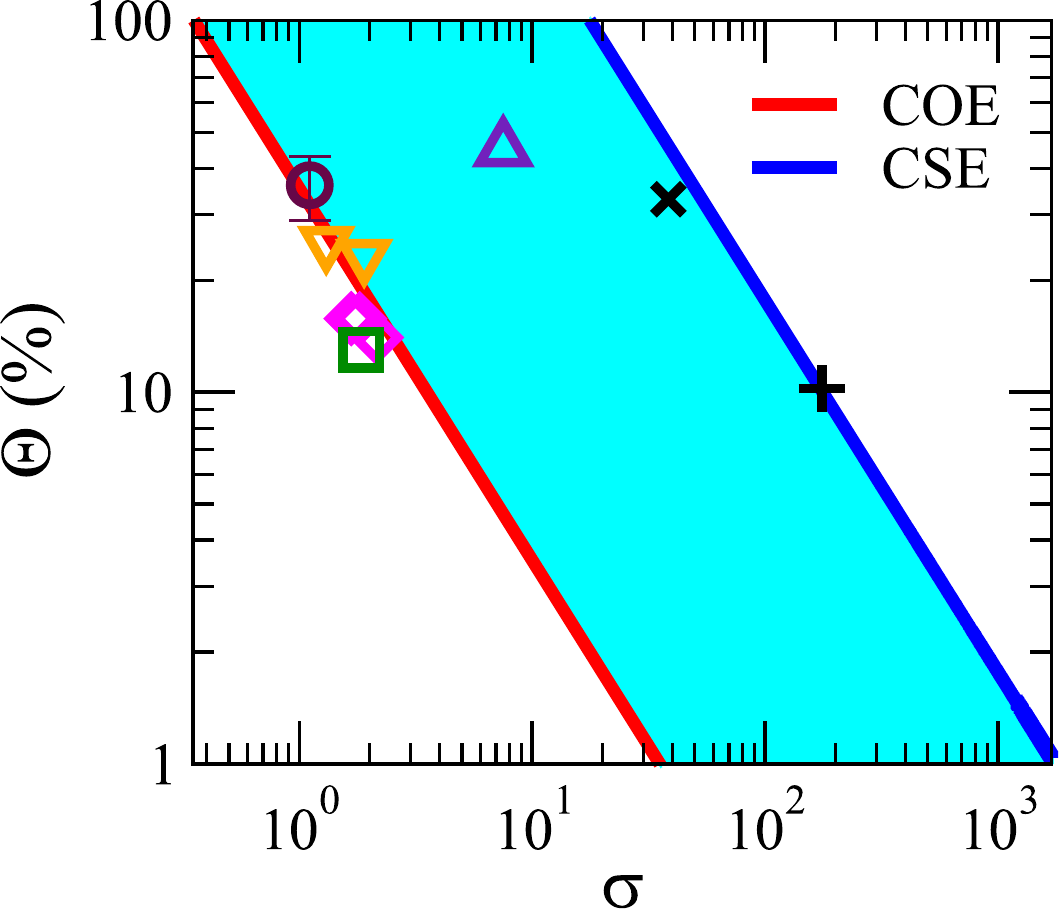}
\caption{The figure shows the $\Theta_{\text{OH}} (\%)$ and $\Theta_{\text{SH}} (\%)$ as a function of dimensionless conductivity $\sigma$. The circle symbol is experimental data of OHA for Ti \cite{https://doi.org/10.48550/arxiv.2109.14847}. The square and triangle up symbols are experimental data of OHA for Ti and W \cite{https://doi.org/10.48550/arxiv.2202.13896}, respectively.  The triangles down and diamonds symbols are experimental data of $\Theta_{\text{LS}} (\%)$ for Cr and Pt \cite{PhysRevResearch.4.033037}, respectively. The plus and times symbols are experimental data of SHA for Pt \cite{PhysRevB.94.060412} and W \cite{doi:10.1063/1.4753947}, respectively. The lines are  the Eq.(\ref{Tc}).}\label{expprb}
\end{figure}

\section{Orbital Hall angle} 

Motivated by recent experiments of OHA \cite{https://doi.org/10.48550/arxiv.2109.14847,https://doi.org/10.48550/arxiv.2202.13896,PhysRevResearch.4.033037} and by the fact that the OHCF are unassailable experimentally, {we use Eq. (\ref{IO}) to obtain two relations that characterize the OHA.} The OHA is defined as the ratio between OHC and charge current
\begin{eqnarray}
\Theta_{\text{OH}}=\frac{I^{o}}{I^c}.\label{II}
\end{eqnarray}
To compute the average of Eq. (\ref{II}), we assume the experimental regime $N\gg1$. Therefore, we have to resort to the CLT and expand (\ref{II}) in the function of $N$ \cite{PhysRevB.102.041107,doi:10.1063/5.0107212}. The average of Eq. (\ref{II}) can be approximated by
\begin{equation}
\left\langle\Theta_{\text{OH}}\right\rangle = \bigg\langle\frac{I^o}{I^c}\bigg\rangle \approx \frac{\langle{I^o}\rangle}
{\langle{I^c}\rangle}.\label{exp}
\end{equation}
By substituting Eq. (\ref{mean}) in Eq. (\ref{exp}), we conclude that
\begin{equation}
\left\langle\Theta_{\text{OH}}\right\rangle = 0, \label{TM}
\end{equation}
for COE and CSE. An equivalent result was obtained for SHA \cite{PhysRevB.102.041107,doi:10.1063/5.0107212}, $\left\langle\Theta_{\text{SH}}\right\rangle = 0$. Although the average of OHA is null, the OHA is expected to have large mesoscopic fluctuations because of its direct dependence on the OHC. 
By following the same methodology that was applied to Eqs. (\ref{IO}) and (\ref{exp}), we can show that
\begin{eqnarray}
\textbf{rms}[\Theta_{\text{OH}}] &=& \frac{\textbf{rms}[I^{o}]}
{\langle{I^c}\rangle}. \label{Tm}
\end{eqnarray}
From Eq.(\ref{Tm}), we can infer the OHA deviation with the knowledge the OHC deviation and the charge current average. 
The former is given by Eq.(\ref{IO}), while the latter is appropriately described by the relation \cite{RevModPhys.69.731,Mello,PhysRevB.61.4453}
\begin{eqnarray}
\langle{{{I^c}}}\rangle = \frac{e^2V}{h}\sigma,\label{Ic2}
\end{eqnarray}
where $\sigma$ is the longitudinal dimensionless conductivity, $\sigma=Nl_e/L$ with $L\gg l_e$.
By substituting Eqs.(\ref{IO}) and (\ref{Ic2}) in (\ref{Tm}), we can infer that the maximum OHA deviation is given by 
\begin{eqnarray}
\Theta_{\text{OH}} \times \sigma = \left\{\begin{array}{cc}
 0.36 & \text{for COE}\\
 0.18 & \text{for CSE}
 \end{array}\right.. \label{Tc}
\end{eqnarray}
{Eq. (\ref{Tc}) is the second main outcome of this work. This shows that the product between maximum OHA deviation $\Theta_{\text{OH}} $ and longitudinal dimensionless conductivity $\sigma$ holds two relationships, which only depend on if the mesoscopic device is a light metal (COE) or a heavy metal (CSE), in contrast with the maximum SHA deviation that holds one relation for heavy metal $\Theta_{\text{SH}} \times \sigma = 0.18$ \cite{PhysRevB.102.041107,doi:10.1063/5.0107212}.}

\section{Comparison with experiments} 
To confirm the validity of Eq. (\ref{Tc}), we compare it with the recent experimental results of Refs. [\onlinecite{https://doi.org/10.48550/arxiv.2109.14847,https://doi.org/10.48550/arxiv.2202.13896,PhysRevResearch.4.033037}]. 

Figure (\ref{expprb}) shows $\Theta_{\text{OH}} (\%)$ as a function of $\sigma$, where the $\sigma$ axis is conveniently normalised as $\sigma=\sigma_{\text{exp}}(\Omega^{-1}\cdot \text{cm}^{-1})/10^{4}(\Omega^{-1}\cdot \text{cm}^{-1})$. 
{The lines are the relations for COE and CSE of Eq. (\ref{Tc}).}
The cyan area is the {\it crossover region} (intermediate SOC) between COE (weak SOC) to CSE (strong SOC).

The circular symbol is experimental data of OHA from Ref. [\onlinecite{https://doi.org/10.48550/arxiv.2109.14847}], which measured the OHE in a light metal Ti. The light metal has weak SOC and, therefore, {follows the COE relation.} The square and triangle up symbols are experimental data of OHA for a light metal Ti and a heavy metal W, respectively, from Ref. [\onlinecite{https://doi.org/10.48550/arxiv.2202.13896}]. Light metal Ti follows the COE, while heavy metal W crossover from COE to CSE. 

The triangles down and diamonds symbols of Fig.(\ref{expprb}) are experimental data of spin-orbital Hall angle $\Theta_{\text{LS}} (\%)$ from [\onlinecite{PhysRevResearch.4.033037}] for light metal Cr and heavy metal Pt, respectively. {They follow the COE relation,} which is expected to be valid for light metals and indicates a pure OHE. The experimental data of conductivity ($\sigma=1/\rho$) and spin-orbital Hall angle ($\Theta_{LS}$) were taken from Fig. 9 of Ref. [\onlinecite{PhysRevResearch.4.033037}] for Cr (samples Cr(9)/Tb(3)/Co(2) and Cr(9)/Gd(3)/Co(2)) and Pt (samples Pt(5)/Co, Pt(5)/Co(2)/Gd(4), and Pt(5)/Co(2)/Tb(4)). 

Furthermore, the plus and times symbols are experimental data of $\Theta_{\text{SH}} (\%)$ for heavy metals Pt [\onlinecite{PhysRevB.94.060412}], and W [\onlinecite{doi:10.1063/1.4753947}], respectively, which follows the CSE [\onlinecite{PhysRevB.102.041107,doi:10.1063/5.0107212}]. 

\begin{figure*}
\centering
\includegraphics[scale = 0.6]{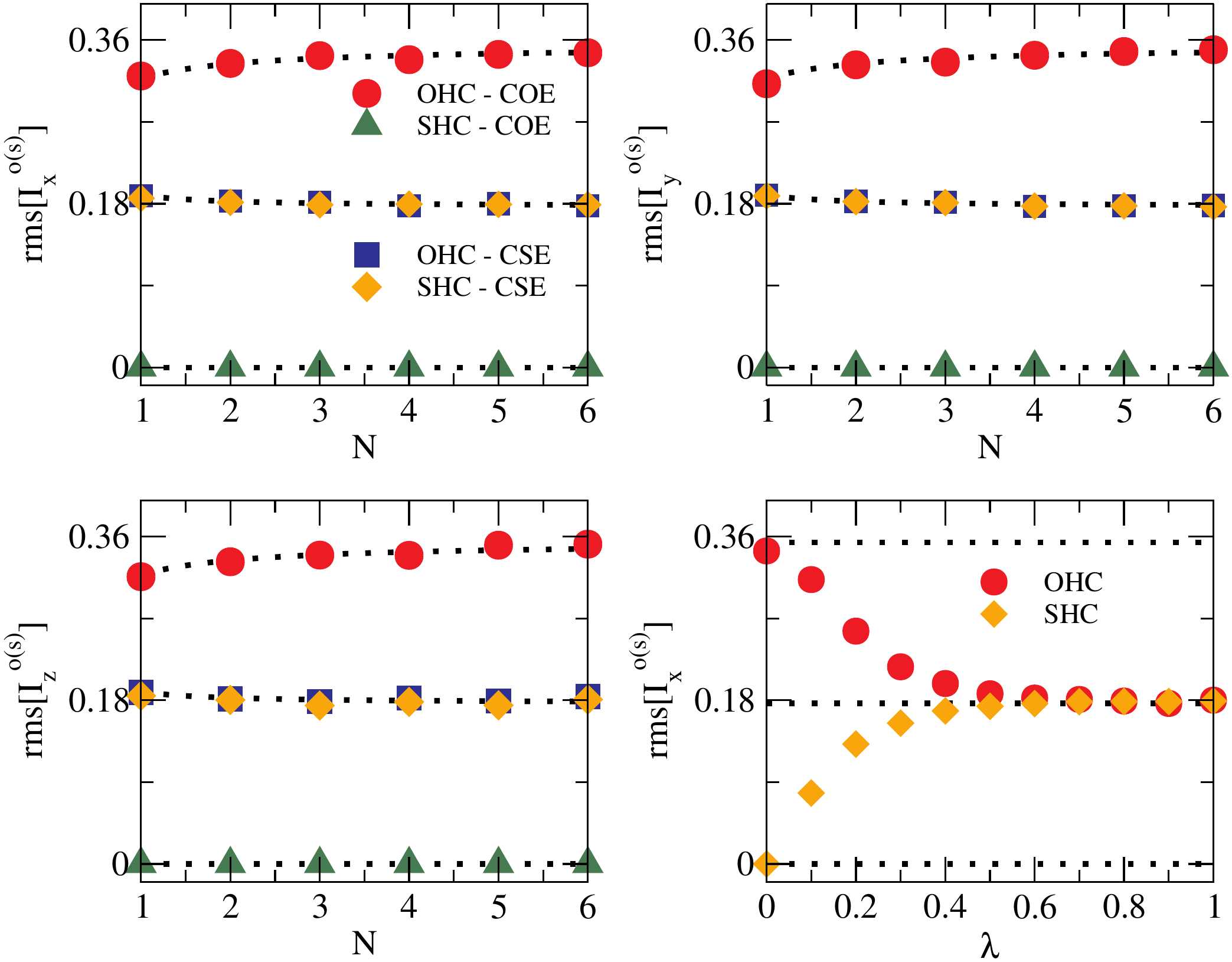}
 \caption{The deviation of OHC and SHC as a function of the thickness of device $N$ for (a) $x$, (b) $y$, and (c) $z$ directions. COE means $\lambda=0.0$ and CSE $\lambda=1.0$, while the dashed lines are Eq. (\ref{IOSS}). (d) The deviation of OHC and SHC as a function SOC parameter $\lambda$ for $N=6$. It shows a crossover between COE  and CSE. The dashed lines are Eq. (\ref{IO}). }\label{figura1}
\end{figure*}

\section{Numerical results}

In this section, we developed two independent numerical calculations of OHCF (SHCF) to confirm Eq. (\ref{IO}) and consequently Eq. (\ref{Tc}). The first is based on the RMT, known as Mahaux-Weidenm\"uller approach \cite{VERBAARSCHOT1985367}, while the second is on the nearest-neighbor tight-binding model  \cite{PhysRevLett.121.086602,PhysRevResearch.2.013177}.

\subsection{Numerical scattering matrix model}

To confirm the analytical results of Eq. (\ref{IO}), we develop an independent numerical calculation based on the Mahaux-Weidenm\"uller approach \cite{VERBAARSCHOT1985367}, a Hamiltonian approach to the random scattering matrix. In this model, the random scattering matrix of a mesoscopic ballistic chaotic device connected to four terminals is given by \cite{PhysRevB.86.235112,PhysRevB.80.125320}
\begin{eqnarray}
\mathcal{S} = \mathbb{1} + 2 i \pi \mathcal{W}^\dagger\left(\mathcal{H}-i\pi\mathcal{W}\mathcal{W}^\dagger\right)^{-1}\mathcal{W}
\label{Sc}
\end{eqnarray}
with dimension $32N \times 32N$, which is described by COE (no SOI) or CSE (with SOI) in the RMT formalism \cite{RevModPhys.69.731}. The Hamiltonian of mesoscopic device 
\begin{eqnarray}
\mathcal{H}&=&\mathcal{H}^0 \otimes \sigma^0 \nonumber\\
&+& \lambda \left(\mathcal{H}^1\otimes l^x \otimes \sigma^x+\mathcal{H}^2\otimes l^y \otimes \sigma^y+\mathcal{H}^3\otimes l^z \otimes \sigma^z\right)\nonumber
\end{eqnarray}
has dimension $8M\times 8M$, while $\mathcal{H}^0$ and $\mathcal{H}^k$ are real symmetric matrices with dimensions $4M\times 4M$ and $M\times M$, respectively. The SOC parameter $\lambda$ ranges between 0 and 1, and $M$ is the number of energy levels in the mesoscopic device. {The RMT regime is gotten when $M\gg N$ \cite{VERBAARSCHOT1985367},} and the Hamiltonian is described by Gaussian Orthogonal Ensemble (GOE) if $\lambda=0$ (no SOC) or Gaussian Symplectic Ensemble (GSE) if $\lambda=1$ (with SOC). Therefore, the Hamiltonian elements follow a Gaussian distribution with zero means \cite{PhysRevB.86.235112,PhysRevB.80.125320}
$$ \langle  \mathcal{H}_{ij}^k \rangle = 0 $$
and variance
$$
\langle  \left(\mathcal{H}_{ij}^k\right)^2\rangle= \left\{\begin{array}{cc}
\frac{2\alpha^2}{M}, & k=0\quad\text{and}\quad i=j\\
\frac{\alpha^2}{M}, & k = 0\quad\text{and}\quad i\neq j\\
\frac{\alpha^2}{4M}, & k\neq 0
\end{array},\right.
$$ where $\alpha=M\Delta/\pi$ is a numerical parameter related to the average spacing, $\Delta$, and energy levels $M$.
Furthermore, $\mathcal{W}=\left(W_1,W_2,W_3,W_4\right)\otimes l^0 \otimes \sigma^0$ is a deterministic matrix with dimension $8M \times 32N$, which connects the $M$ energy levels of devices with $N$ propagating wave modes in the terminals. The coupling matrix $W_p$ has dimension $M\times N$ and satisfies the orthogonality constraint $W_p^\dagger W_q=\frac{\alpha}{\pi}\delta_{ij}$
and its elements are given by \cite{PhysRevB.80.125320}
$$
\left(W_p\right)_{m,n}=\sqrt{\frac{2\alpha}{\pi\left(M+1\right)}}\sin\left[\frac{m\left[\left(p-1\right)N+n\right]\pi}{M+1}\right].
$$

\begin{figure}
\centering
\includegraphics[scale = 0.6]{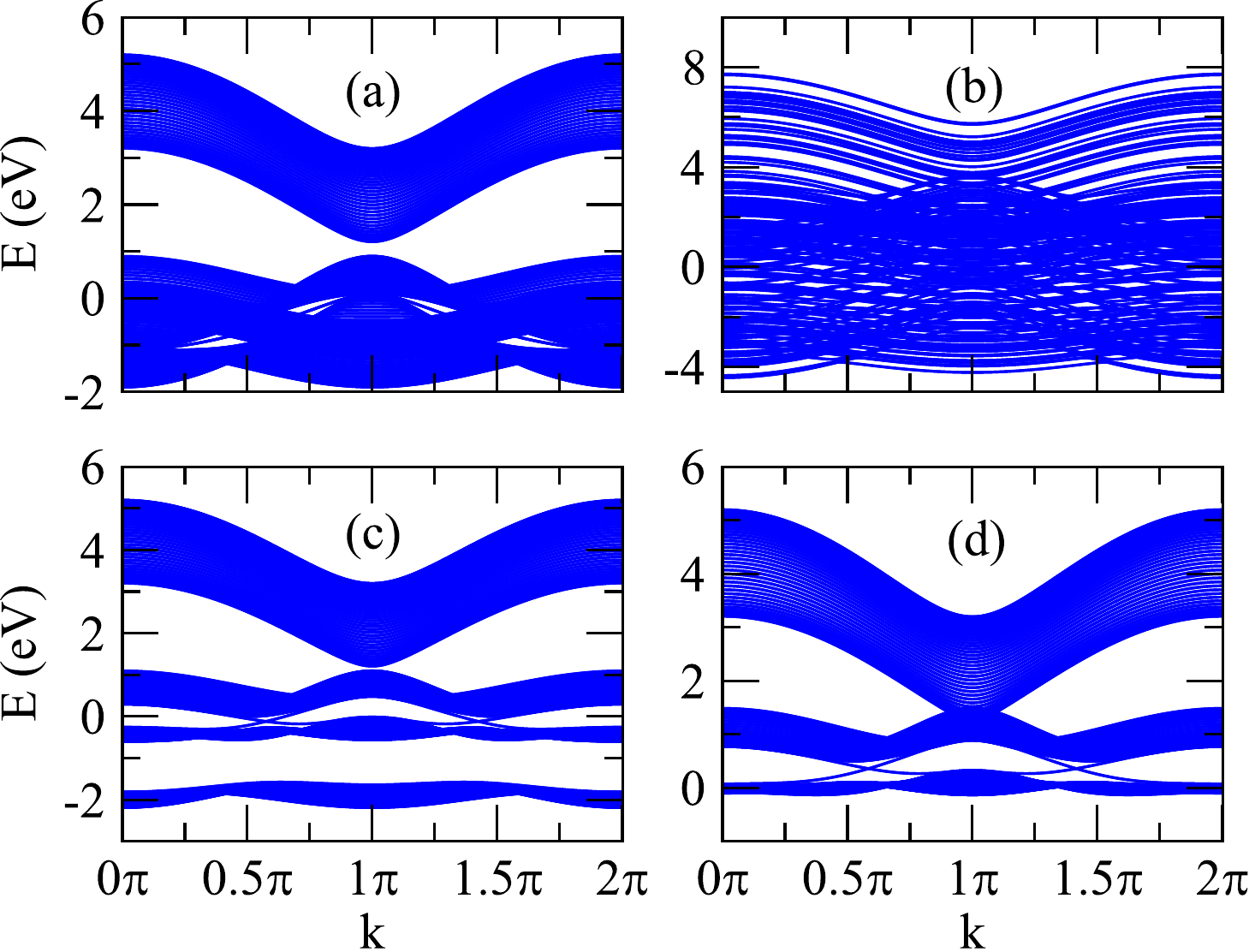}
 \caption{(a) The band structure of the pristine sample, $\lambda=0$ and $U=0$. (b) The one with $U=3$ and $\lambda=0$. The one with $U=0$ and (c) $\lambda=0.5$, and (d) $\lambda=1$.}\label{figura2}
\end{figure}

We developed a numerical calculation of RMT Eq. (\ref{Sc}) and substituted it in Eq.  (\ref{Is}) to calculate the deviation of OHC and SHC as a function of $N$ and $\lambda$. Figure (\ref{figura1}) shows the numerical scattering matrix results for an ensemble with 10000 realizations and $M=200$. 

Figures (\ref{figura1}.a), (\ref{figura1}.b) and (\ref{figura1}.c) show the numeric calculations data (symbols) for the deviation of OHC and SHC $\text{rms}[I^{o(s)}]$ as a function of $N$ for $x, y$, and $z$ directions, respectively. The dashed lines are obtained from Eq. (\ref{IOSS}) and agree with numeric calculations data. { Furthermore, the figures prove that the directions are equivalents and that the results converge to values of Eq. (\ref{IO}).} More specifically, the deviation of OHC converges to 0.36 when $\lambda=0$ (COE) and 0.18 when $\lambda=1$ (CSE), with an increase of $N$ in accord with Eq. (\ref{IO}). Conversely, the deviation of SHC converges to 0.0 when $\lambda=0$ (COE) and 0.18 when $\lambda=1$ (CSE), with an increase of $N$ in accord with [\onlinecite{PhysRevLett.97.066603}]. 

Finally, figure (\ref{figura1}.d) shows the deviation of OHC and SHC as a function of SOC parameter $\lambda$ for $N=6$, which indicates that a crossover between COE and CSE increases $\lambda$ from 0 to 1. More specifically, OHC deviation crossover from 0.36 to 0.18 with an increase of $\lambda$, while SHC deviation crossover from 0.0 to 0.18. Figure (\ref{figura1}) confirms our analytical results of Eqs. (\ref{IOSS}) and (\ref{IO}). 

\begin{figure}
\includegraphics[scale = 0.6]{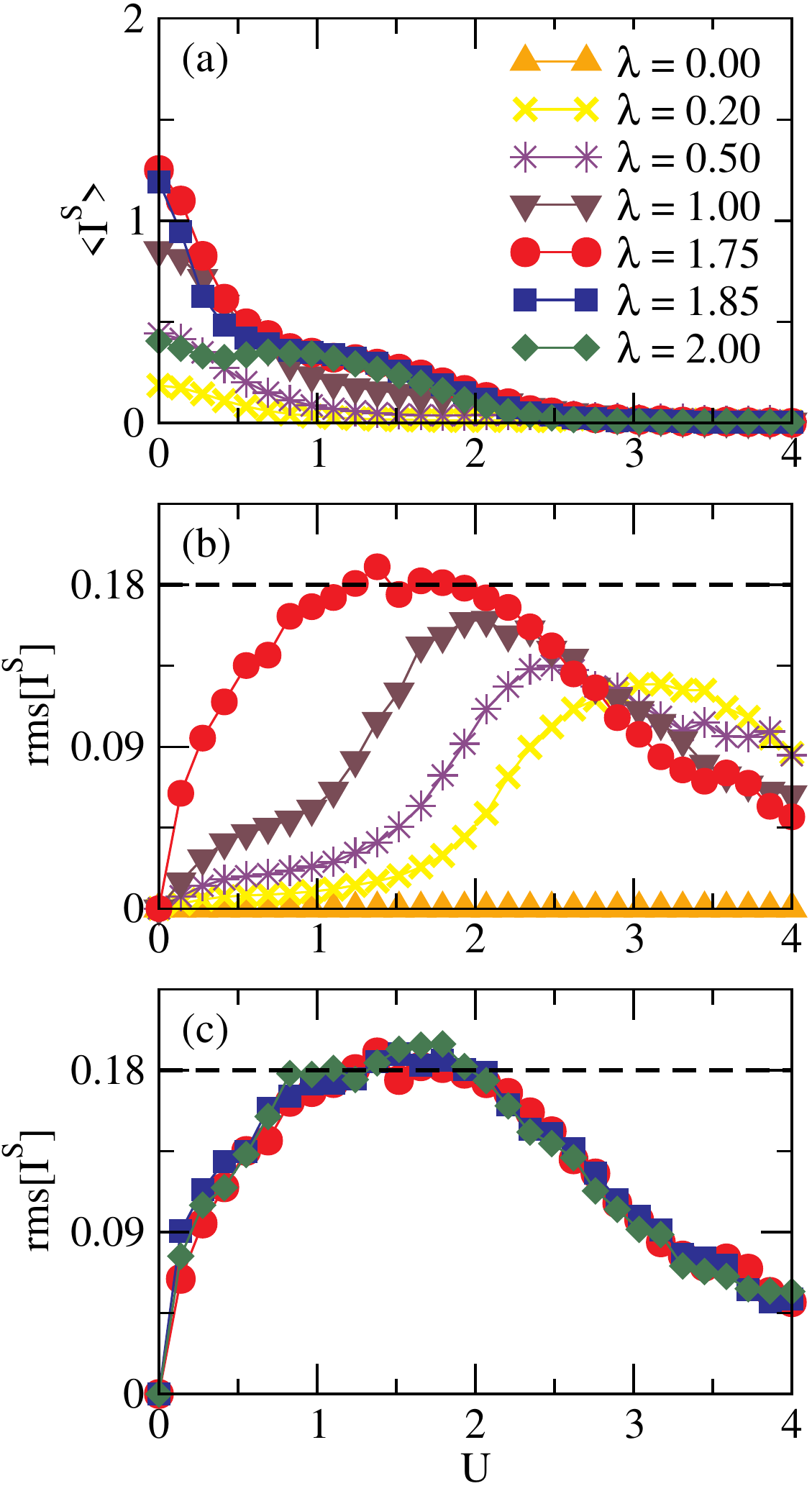}
\caption{(a) The SHC average (in $e^2V/h$) as a function of disorder strength $U$ with different SOC strength $\lambda$ for fixing Fermi energy $E=2.15$. (b) The SHC deviations as a function of $U$ when the SOC parameter $\lambda$ is increased between 0.0 and 1.75. The figure shows a crossover between COE ($\text{rms}[I^s]=0.0$) and CSE ($\text{rms}[I^s]=0.18$) in agreement with Fig. (\ref{figura1}.d).  (c) The SHC deviations as a function of $U$ when the SOC parameter $\lambda$ is increased between 1.75 and 2.0. { The $\text{rms}[I^s]$ is independent of $\lambda$ indicate the reach of the RMT regime.}}\label{figIs}
\end{figure}

\subsection{Tight-binding model}

As a second independent numeric calculation, we study a mesoscopic diffusive device using a two-dimensional square lattice device with a momentum-space orbital texture designed as shown in Fig.(\ref{sample}), in which the nearest-neighbor tight-binding model  \cite{PhysRevLett.121.086602,PhysRevResearch.2.013177} models the lattice with four orbitals (i.e., the $s$ and $p$ orbitals) on each atom.
The Hamiltonian is given by \cite{PhysRevB.103.085113}
\begin{eqnarray}
 H&=&\left(\sum_{\langle i,j\rangle\alpha\beta\sigma}t_{i\alpha,j\beta}c_{i\alpha\sigma}^{\dagger} c_{j\beta\sigma}+\text{H.c.}\right)\nonumber\\
 &+& \sum_{i\alpha\sigma} \left(E_{i\alpha\sigma}+\epsilon_{i\alpha\sigma}\right) c_{i\alpha\sigma}^\dagger c_{i\alpha\sigma}\nonumber\\&+& \lambda \sum_{i\alpha\beta\sigma\delta}\sum_\gamma c_{i\alpha\sigma}^{\dagger}L^\gamma_{\alpha\beta}S^\gamma_{\sigma\delta}c_{i\beta\delta}, \label{TBH}
\end{eqnarray}
where $\{i,j\}$, $\{\alpha,\beta\}$, and $\{\sigma,\delta\}$ are the unit cell, orbital, and spin indices, respectively, and $\gamma = \{x, y, z\}$. The first term represents the nearest-neighbor interaction, where $c_{i\alpha\sigma}$ ($c_{i\alpha\sigma}^\dagger$) is the annihilation (creation) operators and $t_{i\alpha,j\beta}$ denotes hopping integrals. The second is the on-site energy $E_{i\alpha\sigma}$ and Anderson disorder term $\epsilon_{i\alpha\sigma}$. The disorder is realized by an electrostatic potential $\epsilon_i$, which varies randomly from site to site according to a uniform distribution in the interval $\left(-U,U\right)$, where $U$ is the disorder strength. The last is the SOC, where $\lambda$ is the SOC strength, $\vec{L}$ is the angular momentum and $\vec{S}$ is the spin-$1/2$ operator for the electron.
We take the typical Hamiltonian parameters (in eV) $E_s=3.2$, $E_{p_x}=E_{p_y}=E_{p_z}=-0.5$ for on-site energies, $t_s=0.5$, $t_{p\sigma}=0.5$, $t_{p\pi}=0.2$, $t_{sp}=0.5$ for nearest-neighbor hopping amplitudes \cite{PhysRevLett.121.086602,PhysRevResearch.2.013177}. In this tight-binding model, the $sp$ hopping $t_{sp}$ mediates the $k$-dependent hybridization between $p_x$, $p_y$, and $p_z$ orbitals in eigenstates. That means that if $t_{sp}=0$, the orbital texture disappears; hence OHE disappears \cite{PhysRevLett.121.086602}. The two-dimensional square lattice device has a width and length equal to $W = L = 40a$, where $a$ is the square lattice constant.
The numerical calculations were implemented in the KWANT software \cite{kwant}. We used 2000 disorder realization for calculations in this subsection.

Figure (\ref{figura2}.a) shows the band structure of the pristine sample, $\lambda=0$ and $U=0$. The up band is the $s$-character band, while down bans are $p$-character bands. {To ensure we are in the RMT regime $N\gg1$, we use the energy range between 1.5 and 2.5 to calculate OHC.}
Figure (\ref{figura2}.b) shows band structure submitted to disorder strength $U\approx 3$ with $\lambda=0$, while figures (\ref{figura2}.c) and (\ref{figura2}.d) show the one for SOC strength $\lambda=0.5$ and $\lambda=1$ with $U=0$, respectively.

\begin{figure}
\includegraphics[scale = 0.65]{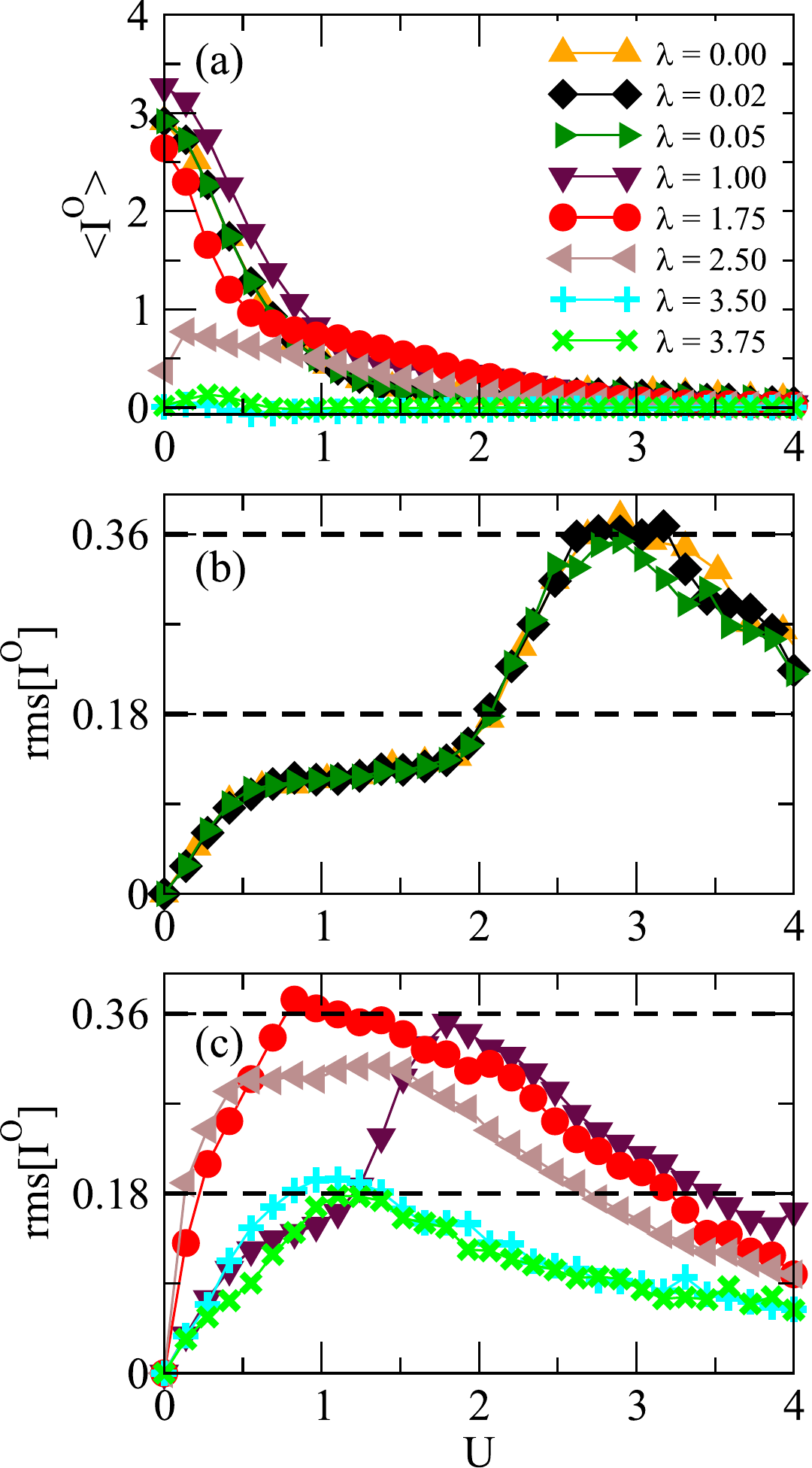}
\caption{The OHC average (in $e^2V/h$) as a function of disorder strength $U$ with different SOC strength $\lambda$ for fixing Fermi energy $E=2.15$. (b) The OHC deviations as a function of $U$ when the SOC parameter $\lambda$ is increased between 0.0, 0.2 and 0.05. {The $\text{rms}[I^s]$ is independent of $\lambda$ indicate the reach of the RMT regime.} (c) The OHC deviations as a function of $U$ when the SOC parameter $\lambda$ is increased between 1.0 and 3.75. The figure shows a crossover between COE ($\text{rms}[I^s]=0.36$) and CSE ($\text{rms}[I^s]=0.18$) in agreement with Fig. (\ref{figura1}.d).  The dashed lines are Eq.(\ref{IO}).}\label{figIO}
\end{figure}

\begin{figure}
\includegraphics[scale = 0.48]{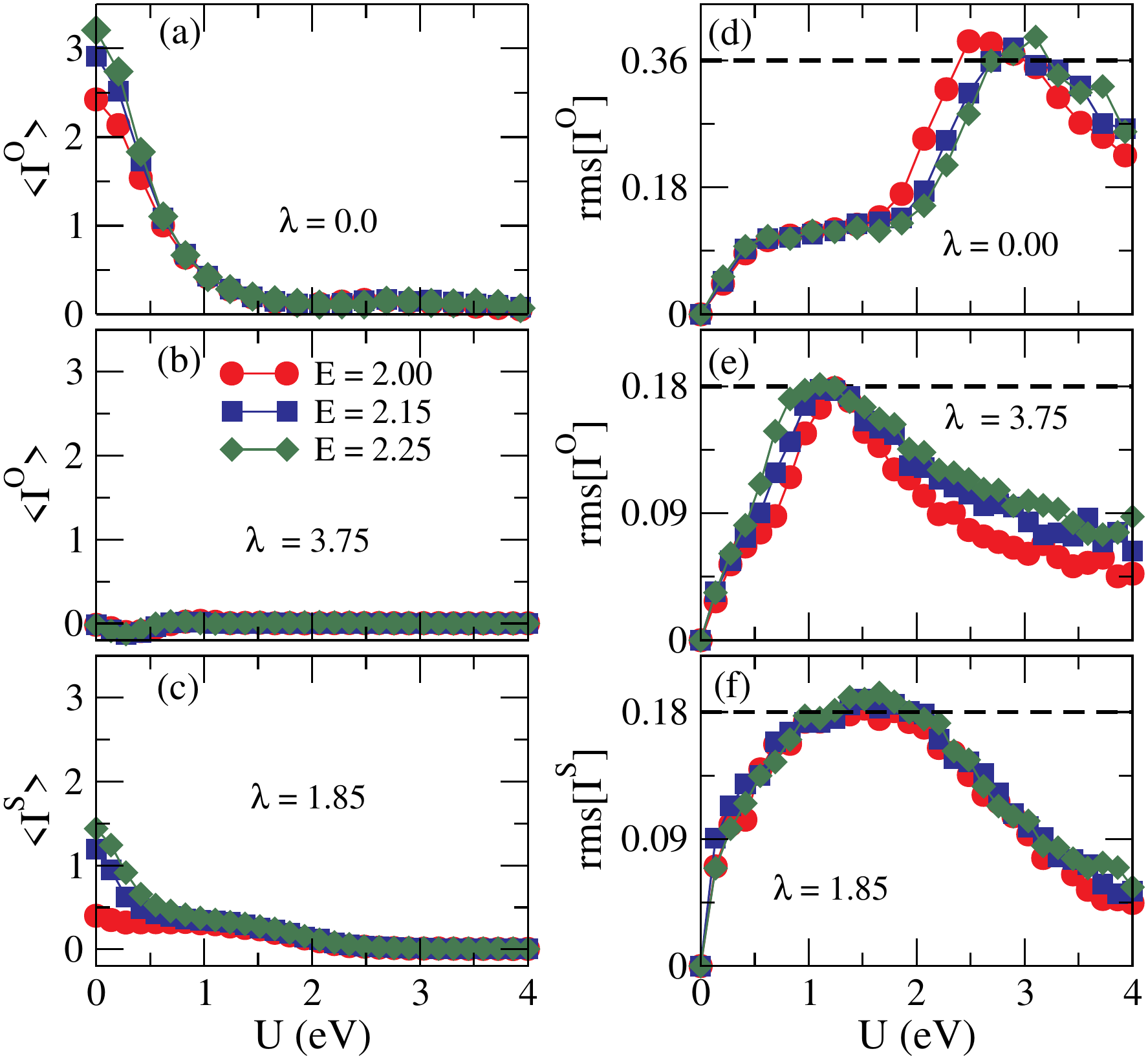}
\caption{The OHC average (in $e^2V/h$) as a function of disorder strength $U$ with fixed SOC strength (a) $\lambda = 0.0$ and (b) $\lambda=3.75$, and (c) SHC average  (in $e^2V/h$) with $\lambda=1.85$ for different Fermi energy. Figures (d), (e), and (f) are their respective deviations as a function of $U$. The dashed lines are Eq.(\ref{IO}).}\label{figIc}
\end{figure}

{Let's start by analyzing the SHCF from Eq. (\ref{Is}) with $\eta=z$ to recover the results of Ref. [\onlinecite{PhysRevLett.97.066603}] to confirm the valid of tight-banding model.} Figure (\ref{figIs}.a) shows the SHC average $\langle{I^s}\rangle$  (in $e^2V/h$) as a function of disorder strength $U$ for a fixed Fermi energy $E=2.15$ and different values of  SOC strength $\lambda$. As expected, for $\lambda=0$, the SHC average is always null, while for $\lambda > 0$, it decreases with increases of $U$.

The SHC deviation $\text{rms}[I^s]$ is shown in Fig. (\ref{figIs}.b) as a function of $U$. The maximum SHC deviation is null for $\lambda=0.0$ (COE) and increases with SOC strength $\lambda=0.2,0.5,1.0$, and 1.75 reaches {the RMT regime of 0.18 (CSE).}  Fig. (\ref{figIs}.b) shows a crossover between COE ($\text{rms}[I^s]=0.0$) and CSE ($\text{rms}[I^s]=0.18$) in agreement with Fig. (\ref{figura1}.d).

On the other side, Fig. (\ref{figIs}.c) shows SHC deviation $\text{rms}[I^s]$ as a function of $U$ for different SOC strength $\lambda=1.75, 1.85$, and 2.0. {In this limit of strong SOC, the SHC deviation becomes independent of SOC strength $\lambda$, indicating the reach of the RMT regime when the disorder strength is $U\approx 1.5$. This disorder does the device satisfy that $\tau_{dwell} \gg \tau_{erg}$.} The results of Fig. (\ref{figIs}) are in agreement with Ref. [\onlinecite{PhysRevLett.97.066603}] and numeric calculations via RMT shown in Fig. (\ref{figura1}). {After the numeric results reach the RMT regime as a function of disorder $U$, the disorder induces a metal-insulator transition in the mesoscopic diffusive device, known as the Anderson transition \cite{RevModPhys.69.731,Mello,PhysRevB.61.4453}. This explains why the numeric result decreases for large disorder $U$ and does not saturate for the RMT regime. The mesoscopic diffusive device behaves as an insulator for large enough disorders.}

{After recovering the results of SHCF,} we are ready to analyze the OHCF using Eq. (\ref{Is}). Figure (\ref{figIO}.a) shows the OHC average $\langle{I^o}\rangle$  (in $e^2V/h$) as a function of $U$ for $E=2.15$ and different values of SOC parameter $\lambda$. When $\lambda=0$, the OHC average is not null and decreases with increases of $U$, in contrast with SHC average  Fig. (\ref{figIs}.a). Furthermore, the OHC average decreases with increases of $\lambda$, in agreement with [\onlinecite{PhysRevLett.121.086602}].

However, we are interested in the amplitude of OHCF. The OHC deviation $\text{rms}[I^o]$ is shown in Fig. (\ref{figIO}.b)  as a function of $U$. {For null SOC strength, $\lambda=0$, the maximum OHC deviation reaches the RMT regime of 0.36, which confirms Eq. (\ref{IO}) for light metals (COE) and agrees with the numeric calculation via RMT shown in Fig. (\ref{figura1}). If we increase the SOC strength $\lambda=0.02$, and 0.05, the maximum OHC deviation remains  0.36.}

{Figure (\ref{figIO}.c) shows the OHC deviation $\text{rms}[I^o]$ as a function of $U$ for strong SOC values. The maximum OHC deviation crossover from 0.36 (COE) to 0.18 (CSE), thus confirming Eq. (\ref{IO}) for heavy metals (CSE).} Furthermore, this result agrees with the numeric simulation via RMT shown in Fig. (\ref{figura1}.d). This behavior can help understudy why the experimental data is on the {\it crossover region} of Fig. (\ref{expprb}). The {\it crossover region} indicates that the SHE and OHE can happen together, and their quantitative contributions cannot be disentangled \cite{PhysRevResearch.4.033037}.
This also is consistent with the interpretation that the OHC is efficiently converted to SHC \cite{Lee2021}.

To confirm the robustness of our results, we fixed the SOC strength and changed the Fermi energy. Figures (\ref{figIc}.a) and (\ref{figIc}.b) show the OHC average as a function of $U$ for $\lambda=0$ and $\lambda=3.75$, respectively, while  Fig. (\ref{figIc}.c) shows the SHC average with $\lambda=1.85$ for different values of Fermi energy. Their respective deviations are shown in Figs. (\ref{figIc}.d), (\ref{figIc}.e), and (\ref{figIc}.f). { For light metal $\lambda=0$ (\ref{figIc}.d), the maximum OHC deviation reaches the RMT regime of 0.36 (COE) independent of Fermi energy. In contrast, for heavy metal $\lambda=3.75$ (\ref{figIc}.d), the one reaches the RMT regime of 0.18 (CSE), which confirms that OHC exhibits mesoscopic fluctuations with amplitudes given by Eq. (\ref{IO}).} Finally, Fig. (\ref{figIc}.f) shows that the maximum SHC deviation reaches 0.18 \cite{PhysRevLett.97.066603}.

\section{Conclusions} 

{We have demonstrated that the OHC exhibits mesoscopic fluctuations. The OHCF displays two amplitudes of 0.36 and 0.18 for light (COE) and heavy (CSE) metals, Eq. (\ref{IO}), respectively; in contrast to the SHCF, which displays one amplitude of 0.18 for heavy metals (CSE). From the view of RMT, there is a crossover from COE to CSE when the SOC is increased, as shown in Fig. (\ref{figura1}.d). In other words, we have a pure OHE when the amplitude is 0.36 (COE), and we can have OHE and SHE happen together when the amplitude is 0.18 (CSE), as shown in Fig. (\ref{expprb}). Furthermore, the OHCF leads to two relationships between the maximum OHA deviation and the dimensionless conductivity $\sigma$ given by Eq. (\ref{Tc}). The two relationships are in agreement with the experimental data of [\onlinecite{https://doi.org/10.48550/arxiv.2109.14847,https://doi.org/10.48550/arxiv.2202.13896,PhysRevResearch.4.033037}], Fig. (\ref{expprb}).} 

{The results are calculated analytically via RMT and supported by numerical calculations based on the tight-binding model.  They are valid for ballistic chaotic and mesoscopic diffusive devices in the limit when the mean dwell time of the electrons is much longer than the time needed for ergodic exploration of the phase space, $\tau_{dwell} \gg \tau_{erg}$.}

{This work brings a new perspective on OHE and may help to give a deeper understanding of the effect. 
Furthermore, similar to what happens with SHCF \cite{PhysRevLett.101.016804,doi:10.1063/5.0107212,PhysRevB.93.115120}, we expect that OHCF in topological insulators \cite{PhysRevB.101.075429,PhysRevB.102.035409} follow the same amplitudes of Eq. (\ref{IO}) when $N\gg1$.
The presented methodology can be extended to other effects, such as the spin Nernst effect \cite{Meyer2017}, giving rise to a set of relationships, such as Eq. (\ref{Tc}).}

\begin{acknowledgments}
DBF acknowledges a scholarship from Funda\c{c}\~ao de Amparo a Ci\^encia e Tecnologia de Pernambuco (FACEPE, Grant IBPG-0253-1.04/22).
ALRB acknowledges financial support from Conselho Nacional de Desenvolvimento Cient\'{\i}fico e Tecnol\'ogico (CNPq, Grant 309457/2021) and Coordenação de Aperfeiçoamento de Pessoal de Nível Superior (CAPES).
\end{acknowledgments}

\appendix
\section{Landauer-B\"uttiker model}\label{A}

From the Landauer-B\"uttiker model, the transversal orbital (spin) Hall current OHC (SHC) through the $i$th terminal in the linear regime at low temperature is
\begin{equation}
    I^{o(s)}_{i,\eta} = \frac{e^2}{h}\sum_{j} \tau_{ij,\eta}^{o(s)} \left( V_i - V_j \right),\label{IOS2}
\end{equation} 
where the orbital (spin) transmission coefficient is calculated from the transmission and reflection blocks of the scattering matrix
\begin{eqnarray}
     \tau_{ij,\eta}^{o(s)}=\textbf{Tr}\left[\left(\mathcal{S}_{ij}\right)^{\dagger} \mathcal{P}^{o(s)}_\eta \mathcal{S}_{ij}\right], \;         \mathcal{S} = \left[ \begin{array}{cccc}
    r_{11} & t_{12} & t_{13} & t_{14}  \\
    t_{21} & r_{22} & t_{23} & t_{24} \\
    t_{31} & t_{32} & r_{33} & t_{34} \\
    t_{41} & t_{42} & t_{43} & r_{44} 
\end{array} \right]. \label{SC}
\end{eqnarray}
The scattering matrix $\mathcal{S}$ has dimension $32N \times 32N$, while its blocks $\mathcal{S}_{ij}$ have dimension $8N \times 8N$, where $N$ is the number of  propagating wave modes in the terminals. The matrix $\mathcal{P}^{o(s)}_\eta = \mathbb{1}_N \otimes l^\eta \otimes \sigma^0$ $\left(\mathbb{1}_N \otimes l^0 \otimes \sigma^\eta\right)$ is a orbital (spin) projector with dimension $8N\times 8N$, and $\mathbb{1}_N$ is a identity matrix $N\times N$. The index $\eta = \{0,x,y,z\}$, while $$l^0=(l^\eta)^2, \quad \text{and} \quad \sigma^0=(\sigma^\eta)^2.$$ The $l^\eta$ matrix is the orbital angular momentum matrices
\begin{eqnarray}
    l^x &=& \left[ \begin{array}{cccc}
    0 & 0 & 0 & 0 \\
    0 & 0 & 0 & 0 \\
    0 & 0 & 0 & -i \\
    0 & 0 & i & 0
\end{array} \right],\nonumber\\
\quad l^y &=& \left[ \begin{array}{cccc}
    0 & 0 & 0 & 0 \\
    0 & 0 & 0 & i \\
    0 & 0 & 0 & 0 \\
    0 & -i & 0 & 0
\end{array} \right],\nonumber\\
\quad l^z &=& \left[ \begin{array}{cccc}
    0 & 0 & 0 & 0 \\
    0 & 0 & -i & 0 \\
    0 & i & 0 & 0 \\
    0 & 0 & 0 & 0
\end{array} \right],
\end{eqnarray}
and $\sigma^\eta$ is Pauli matrices
\begin{eqnarray}
     \sigma^x = \left[ \begin{array}{cc}
    0 & 1  \\
    1 & 0  
\end{array} \right],
\quad \sigma^y = \left[ \begin{array}{cc}
    0 & -i  \\
    i & 0  
\end{array} \right],
\quad \sigma^z = \left[ \begin{array}{cc}
    1 & 0  \\
    0 & -1  
\end{array} \right].
\end{eqnarray}

The charge current $I^{c}_{i,0}$ is defined by $\eta=0$, while OHC (SHC) $I^{o(s)}_{i,\eta}$ by $\eta=\{x,y,z\}$.
 The mesoscopic device is connected to four semi-infinite terminals submitted to voltages $V_i$, Fig.1 of the main text. From Eq. (\ref{IOS2}), the charge current and OHC (SHC) can be written as
\begin{equation}
    I^{o(s)}_{1,\eta} = \frac{e^2}{h}\left[\tau^{o(s)}_{12,\eta}(V_1 - V_2) + \tau^{o(s)}_{13,\eta}(V_1 - V_3) + \tau^{o(s)}_{14,\eta}(V_1 - V_4)\right], \label{IO_1}
\end{equation}
\begin{equation}
    I^{o(s)}_{2,\eta} = \frac{e^2}{h}\left[ \tau^{o(s)}_{21,\eta}(V_2 - V_1) + \tau^{o(s)}_{23,\eta}(V_2 - V_3) + \tau^{o(s)}_{24,\eta}(V_2 - V_4)\right],\label{IO_2}
\end{equation}
\begin{equation}
   I^{o(s)}_{3,\eta} = \frac{e^2}{h} \left[\tau^{o(s)}_{31,\eta}(V_3 - V_1) + \tau^{o(s)}_{32,\eta}(V_3 - V_2) + \tau^{o(s)}_{34,\eta}(V_3 - V_4)\right], \label{IO_3}
\end{equation}
\begin{equation}
    I^{o(s)}_{4,\eta} = \frac{e^2}{h} \left[\tau^{o(s)}_{41,\eta}(V_4 - V_1) + \tau^{o(s)}_{42,\eta}(V_4 - V_2) + \tau^{o(s)}_{43,\eta}(V_4 - V_3)\right].\label{IO_4}
\end{equation}
Assuming that the charge current is conserved in the longitudinal terminals, $I^c_{1,0}=-I^c_{2,0}=I^c$, we obtain from Eqs. (\ref{IO_1}) and (\ref{IO_2}) the longitudinal charge current \cite{PhysRevB.72.075361,Nikolic_2007,PhysRevLett.98.196601}
\begin{eqnarray}
2I^c&=&\frac{e^2V}{h}\bigg[\frac{1}{2}(8N - \tau_{11,0} - \tau_{22,0} + \tau_{21,0}+\tau_{12,0}) \nonumber\\
    &+& \frac{V_3}{V}(\tau_{23,0}-\tau_{13,0}) + \frac{V_4}{V}(\tau_{24,0}-\tau_{14,0}) \bigg]. \label{22IO}
\end{eqnarray}
Where it was considered $V_1=V/2$, $V_2=-V/2$, being $V$ a constant, and $\sum^4_{j=1} \tau_{ij,\eta}^{o(s)} = 4N\delta_{\eta 0}$.
The OHC (SHC) is obtained from Eqs. (\ref{IO_3}) and (\ref{IO_4}), then 
\begin{eqnarray}
I^{o(s)}_{i,\eta} = \frac{e^2V}{h}\left[\frac{1}{2}\left(\tau_{i2,\eta}^{o(s)}-\tau_{i1,\eta}^{o(s)}\right)
- \tau_{i3,\eta}^{o(s)}\frac{V_3}{V} + \tau_{i4,\eta}^{o(s)}\frac{V_4}{V}\right],
\label{Iss}
\end{eqnarray}
for $i=3,4$ and $\quad\eta = \{x,y,z\}$.
Eq. (\ref{Iss}) is Eq. (\ref{Is}). 
Finally, we assume that the charge current vanishes in the transverse terminals $I^{c}_{3,0}=I^{c}_{4,0}=0$. Then, from Eqs. (\ref{IO_3}) and (\ref{IO_4}) we obtain
\begin{eqnarray}
     \frac{V_i}{V}=\frac{1}{2}\frac{\tau_{ij,0}(\tau_{j2,0}-\tau_{j1,0})+(\tau_{i2,0}-\tau_{i1,0})(4N - \tau_{jj,0})}{(\tau_{43,0}\tau_{34,0})-(4N-\tau_{33,0})(4N-\tau_{44,0})}, \label{VObarra3}
\end{eqnarray}
for $i,j=3,4$ with $i\neq j$.

\section{Random Matrix Theory}\label{B}

Applying the method of Ref. [\onlinecite{doi:10.1063/1.531667}] to Eq.(\ref{SC}), we found for the circular orthogonal ensemble (COE, no SOI) and the circular symplectic ensemble (CSE, with SOI), that the average of orbital transmission coefficients are
\begin{eqnarray}
    \langle \tau_{ij,\eta}^o \rangle_{\text{COE}} &=& \delta_{\eta0}\frac{4N\left(2N +\delta_{ij}\right)}{8N+1}, \label{TCOE}\\ 
    \langle \tau_{ij,\eta}^{o} \rangle_{\text{CSE}} &=& \delta_{\eta0}\frac{4N\left(4N +\delta_{ij}\right)}{16N-1}. \label{TCSE}
\end{eqnarray}
For the second moment of orbital transmission coefficients
\begin{widetext}
\begin{eqnarray}
    \langle \tau_{ij,\eta}^{o}\tau_{kl,\mu}^{o} \rangle_{\text{COE}} &=& \frac{4}{N(8N+1)(8N+3)} \Bigg[  2(4N+1) \delta_{\eta0}\delta_{\mu0}\big[2N^4 + N^3 (\delta_{kl} + \delta_{ij})\big]  \nonumber \\
    &+& (4N+1) \big[N^2 (\delta_{\eta0}\delta_{\mu0}(\delta_{ij} \delta_{kl} + \delta_{il}\delta_{jk}) + \delta_{ik}\delta_{jl} \delta_{\eta\mu})+ N\delta_{iljk}\delta_{\eta\mu}\delta_{\eta0} \big]  \nonumber  \\
    &-& N^3\big[\delta_{\eta0}\delta_{\mu0}(\delta_{jk}+\delta_{jl}+\delta_{il}) + \delta_{ik}\delta_{\eta\mu}\big] - N\delta_{ilkj}\delta_{\eta\mu}\delta_{\eta0} \nonumber  \\
    &-& N^2\delta_{\eta0} \big[ \delta_{\mu0}(\delta_{jlk}+ \delta_{ilj}) + \frac{1}{2}\delta_{\eta\mu}(\delta_{ikj}+\delta_{ilk}+\delta_{ikl} + \delta_{ijkl})\big]  \Bigg],
    \label{TTCOE}\\
    \langle \tau_{ij,\eta}^{o}\tau_{kl,\mu}^{o} \rangle_{\text{CSE}} &=& \frac{1}{N(16N-1)(16N-3)} \Big[  2(8N-1) \big[\delta_{\eta0}\delta_{\mu0}(16N^4 + 4N^3 (\delta_{kl} + \delta_{ij}) + N^2(\delta_{ij}\delta_{kl} + \delta_{il}\delta_{jk}))  \nonumber  \\
    &+&N^2\delta_{ik}\delta_{jl}\delta_{\eta\mu}\big] + 4N^3 \big[\delta_{\eta0}\delta_{\mu0} (\delta_{jk} + \delta_{jl} + \delta_{il})+ \delta_{ik}\delta_{\eta\mu}\big] + N\delta_{\eta\mu}\delta_{\eta0}\big[ \delta_{ilkj} + (8N-1) \delta_{iljk} \big] \nonumber  \\
    &+& N^2\delta_{\eta0}\big[2\delta_{\mu0}(\delta_{jlk}+\delta_{ilj}) + \delta_{\eta\mu} (\delta_{ikj}+\delta_{ilk} +\delta_{ikl} + \delta_{ijkl})\big] \Bigg],
    \label{TTCSE}
\end{eqnarray}
\end{widetext}
where $\eta,\mu=\{0,x,y,z\}$. On the other side, the average of spin transmission coefficients is $\langle \tau_{ij,\eta}^s \rangle_{\text{COE}} = 0$ and $\langle \tau_{ij,\eta}^s \rangle_{\text{CSE}} = \langle \tau_{ij,\eta}^o \rangle_{\text{CSE}} $, and for the second moments are $ \langle \tau_{ij,\eta}^{o}\tau_{kl,\mu}^{o} \rangle_{\text{COE}}=0$ and $\langle \tau_{ij,\eta}^{s}\tau_{kl,\mu}^{s} \rangle_{\text{CSE}}=\langle \tau_{ij,\eta}^{o}\tau_{kl,\mu}^{o} \rangle_{\text{CSE}}$. 

Taking Eqs. (\ref{TCOE}) and (\ref{TCSE}) into account, we conclude that the average of OHC (SHC), Eq. (\ref{Is}), is null
\begin{eqnarray}
\langle I^{o(s)}_{i,\eta} \rangle &=& \frac{e^2V}{h}\left[\frac{1}{2}\left(\langle \tau_{i2,\eta}^{o}\rangle -\langle\tau_{i1,\eta}^{o}\rangle\right)\right.\nonumber\\
&-& \left.\langle\tau_{i3,\eta}^{o}\rangle\frac{\langle V_3\rangle}{V} + \langle\tau_{i4,\eta}^{o}\rangle\frac{\langle V_4\rangle}{V}\right] = 0,
\end{eqnarray}
for COE and CSE, which proves Eq. (\ref{mean}). The transversal potentials $V_{3,4}$ are orbital independent because they depend only on $\tau_{ij,0}$. Thus, they do not have correlation with $\tau_{ij,\eta}^{o}$, see Eq. (\ref{VObarra3}). Furthermore, applying  Eqs. (\ref{TCOE}), (\ref{TCSE}), (\ref{TTCOE}), (\ref{TTCSE}) to Eq.(\ref{VObarra3}), we obtain that $\langle V_{3,4}\rangle = 0$.

Finally, the variance of OHC is defined as $$
\textbf{var}[I^{o}_{i,\eta}]=\left\langle {I^{o}_{i,\eta}}^2\right\rangle-\left\langle I^{o}_{i,\eta}\right\rangle^2 = \left\langle  {I^{o}_{i,\eta}}^2\right\rangle.
$$ Using Eqs. (\ref{TTCOE}), (\ref{TTCSE}), we can obtain that
\begin{eqnarray}
\textbf{var}[I^{o}_{i,\eta}]=\left(\frac{e^2V}{h}\right)^2 \times \Bigg\{\begin{array}{cc}
\frac{2N\left(4N+1\right)}{\left(8N+1\right)\left(8N+3\right)} & \text{for COE}\\
\frac{N\left(8N-1\right)}{\left(16N-1\right)\left(16N-3\right)} & \text{for CSE}
\end{array},\label{IOO}
\end{eqnarray}
for  $i=3,4$ and $\eta = \{x,y,z\}$.
The Eq. (\ref{IOO}) proves the Eq. (\ref{IOSS}) of the main text.

\bibliography{ref}

\begin{thebibliography}{64}
\expandafter\ifx\csname natexlab\endcsname\relax\def\natexlab#1{#1}\fi
\expandafter\ifx\csname bibnamefont\endcsname\relax
  \def\bibnamefont#1{#1}\fi
\expandafter\ifx\csname bibfnamefont\endcsname\relax
  \def\bibfnamefont#1{#1}\fi
\expandafter\ifx\csname citenamefont\endcsname\relax
  \def\citenamefont#1{#1}\fi
\expandafter\ifx\csname url\endcsname\relax
  \def\url#1{\texttt{#1}}\fi
\expandafter\ifx\csname urlprefix\endcsname\relax\def\urlprefix{URL }\fi
\providecommand{\bibinfo}[2]{#2}
\providecommand{\eprint}[2][]{\url{#2}}

\bibitem[{\citenamefont{Dyakonov and Perel}(1971)}]{pereldois}
\bibinfo{author}{\bibfnamefont{M.}~\bibnamefont{Dyakonov}} \bibnamefont{and}
  \bibinfo{author}{\bibfnamefont{V.}~\bibnamefont{Perel}},
  \bibinfo{journal}{Physics Letters A} \textbf{\bibinfo{volume}{A 35}},
  \bibinfo{pages}{459} (\bibinfo{year}{1971}),
  \urlprefix\url{https://doi.org/10.1016/0375-9601(71)90196-4}.

\bibitem[{\citenamefont{Hirsch}(1999)}]{spinhallh}
\bibinfo{author}{\bibfnamefont{J.~E.} \bibnamefont{Hirsch}},
  \bibinfo{journal}{Phys. Rev. Lett.} \textbf{\bibinfo{volume}{83}},
  \bibinfo{pages}{1834} (\bibinfo{year}{1999}),
  \urlprefix\url{https://link.aps.org/doi/10.1103/PhysRevLett.83.1834}.

\bibitem[{\citenamefont{Kato et~al.}(2004)\citenamefont{Kato, Myers, Gossard,
  and Awschalom}}]{Kato1910}
\bibinfo{author}{\bibfnamefont{Y.~K.} \bibnamefont{Kato}},
  \bibinfo{author}{\bibfnamefont{R.~C.} \bibnamefont{Myers}},
  \bibinfo{author}{\bibfnamefont{A.~C.} \bibnamefont{Gossard}},
  \bibnamefont{and} \bibinfo{author}{\bibfnamefont{D.~D.}
  \bibnamefont{Awschalom}}, \bibinfo{journal}{Science}
  \textbf{\bibinfo{volume}{306}}, \bibinfo{pages}{1910} (\bibinfo{year}{2004}),
  ISSN \bibinfo{issn}{0036-8075},
  \urlprefix\url{https://science.sciencemag.org/content/306/5703/1910}.

\bibitem[{\citenamefont{Wunderlich et~al.}(2005)\citenamefont{Wunderlich,
  Kaestner, Sinova, and Jungwirth}}]{PhysRevLett.94.047204}
\bibinfo{author}{\bibfnamefont{J.}~\bibnamefont{Wunderlich}},
  \bibinfo{author}{\bibfnamefont{B.}~\bibnamefont{Kaestner}},
  \bibinfo{author}{\bibfnamefont{J.}~\bibnamefont{Sinova}}, \bibnamefont{and}
  \bibinfo{author}{\bibfnamefont{T.}~\bibnamefont{Jungwirth}},
  \bibinfo{journal}{Phys. Rev. Lett.} \textbf{\bibinfo{volume}{94}},
  \bibinfo{pages}{047204} (\bibinfo{year}{2005}),
  \urlprefix\url{https://link.aps.org/doi/10.1103/PhysRevLett.94.047204}.

\bibitem[{\citenamefont{Nikoli\ifmmode~\acute{c}\else \'{c}\fi{}
  et~al.}(2005)\citenamefont{Nikoli\ifmmode~\acute{c}\else \'{c}\fi{}, Z\^arbo,
  and Souma}}]{PhysRevB.72.075361}
\bibinfo{author}{\bibfnamefont{B.~K.}
  \bibnamefont{Nikoli\ifmmode~\acute{c}\else \'{c}\fi{}}},
  \bibinfo{author}{\bibfnamefont{L.~P.} \bibnamefont{Z\^arbo}},
  \bibnamefont{and} \bibinfo{author}{\bibfnamefont{S.}~\bibnamefont{Souma}},
  \bibinfo{journal}{Phys. Rev. B} \textbf{\bibinfo{volume}{72}},
  \bibinfo{pages}{075361} (\bibinfo{year}{2005}),
  \urlprefix\url{https://link.aps.org/doi/10.1103/PhysRevB.72.075361}.

\bibitem[{\citenamefont{Raimondi et~al.}(2006)\citenamefont{Raimondi, Gorini,
  Schwab, and Dzierzawa}}]{PhysRevB.74.035340}
\bibinfo{author}{\bibfnamefont{R.}~\bibnamefont{Raimondi}},
  \bibinfo{author}{\bibfnamefont{C.}~\bibnamefont{Gorini}},
  \bibinfo{author}{\bibfnamefont{P.}~\bibnamefont{Schwab}}, \bibnamefont{and}
  \bibinfo{author}{\bibfnamefont{M.}~\bibnamefont{Dzierzawa}},
  \bibinfo{journal}{Phys. Rev. B} \textbf{\bibinfo{volume}{74}},
  \bibinfo{pages}{035340} (\bibinfo{year}{2006}),
  \urlprefix\url{https://link.aps.org/doi/10.1103/PhysRevB.74.035340}.

\bibitem[{\citenamefont{Gorini}(2022)}]{GORINI2022}
\bibinfo{author}{\bibfnamefont{C.}~\bibnamefont{Gorini}}, in
  \emph{\bibinfo{booktitle}{Reference Module in Materials Science and Materials
  Engineering}} (\bibinfo{publisher}{Elsevier}, \bibinfo{year}{2022}), ISBN
  \bibinfo{isbn}{978-0-12-803581-8},
  \urlprefix\url{https://www.sciencedirect.com/science/article/pii/B9780323908009001013}.

\bibitem[{\citenamefont{Sinova et~al.}(2015)\citenamefont{Sinova, Valenzuela,
  Wunderlich, Back, and Jungwirth}}]{RevModPhys.87.1213}
\bibinfo{author}{\bibfnamefont{J.}~\bibnamefont{Sinova}},
  \bibinfo{author}{\bibfnamefont{S.~O.} \bibnamefont{Valenzuela}},
  \bibinfo{author}{\bibfnamefont{J.}~\bibnamefont{Wunderlich}},
  \bibinfo{author}{\bibfnamefont{C.~H.} \bibnamefont{Back}}, \bibnamefont{and}
  \bibinfo{author}{\bibfnamefont{T.}~\bibnamefont{Jungwirth}},
  \bibinfo{journal}{Rev. Mod. Phys.} \textbf{\bibinfo{volume}{87}},
  \bibinfo{pages}{1213} (\bibinfo{year}{2015}),
  \urlprefix\url{https://link.aps.org/doi/10.1103/RevModPhys.87.1213}.

\bibitem[{\citenamefont{Avsar et~al.}(2020)\citenamefont{Avsar, Ochoa, Guinea,
  \"Ozyilmaz, van Wees, and Vera-Marun}}]{coloquiumspintronics}
\bibinfo{author}{\bibfnamefont{A.}~\bibnamefont{Avsar}},
  \bibinfo{author}{\bibfnamefont{H.}~\bibnamefont{Ochoa}},
  \bibinfo{author}{\bibfnamefont{F.}~\bibnamefont{Guinea}},
  \bibinfo{author}{\bibfnamefont{B.}~\bibnamefont{\"Ozyilmaz}},
  \bibinfo{author}{\bibfnamefont{B.~J.} \bibnamefont{van Wees}},
  \bibnamefont{and} \bibinfo{author}{\bibfnamefont{I.~J.}
  \bibnamefont{Vera-Marun}}, \bibinfo{journal}{Rev. Mod. Phys.}
  \textbf{\bibinfo{volume}{92}}, \bibinfo{pages}{021003}
  (\bibinfo{year}{2020}),
  \urlprefix\url{https://link.aps.org/doi/10.1103/RevModPhys.92.021003}.

\bibitem[{\citenamefont{Sagasta et~al.}(2016)\citenamefont{Sagasta, Omori,
  Isasa, Gradhand, Hueso, Niimi, Otani, and Casanova}}]{PhysRevB.94.060412}
\bibinfo{author}{\bibfnamefont{E.}~\bibnamefont{Sagasta}},
  \bibinfo{author}{\bibfnamefont{Y.}~\bibnamefont{Omori}},
  \bibinfo{author}{\bibfnamefont{M.}~\bibnamefont{Isasa}},
  \bibinfo{author}{\bibfnamefont{M.}~\bibnamefont{Gradhand}},
  \bibinfo{author}{\bibfnamefont{L.~E.} \bibnamefont{Hueso}},
  \bibinfo{author}{\bibfnamefont{Y.}~\bibnamefont{Niimi}},
  \bibinfo{author}{\bibfnamefont{Y.}~\bibnamefont{Otani}}, \bibnamefont{and}
  \bibinfo{author}{\bibfnamefont{F.}~\bibnamefont{Casanova}},
  \bibinfo{journal}{Phys. Rev. B} \textbf{\bibinfo{volume}{94}},
  \bibinfo{pages}{060412} (\bibinfo{year}{2016}),
  \urlprefix\url{https://link.aps.org/doi/10.1103/PhysRevB.94.060412}.

\bibitem[{\citenamefont{Pai et~al.}(2012)\citenamefont{Pai, Liu, Li, Tseng,
  Ralph, and Buhrman}}]{doi:10.1063/1.4753947}
\bibinfo{author}{\bibfnamefont{C.-F.} \bibnamefont{Pai}},
  \bibinfo{author}{\bibfnamefont{L.}~\bibnamefont{Liu}},
  \bibinfo{author}{\bibfnamefont{Y.}~\bibnamefont{Li}},
  \bibinfo{author}{\bibfnamefont{H.~W.} \bibnamefont{Tseng}},
  \bibinfo{author}{\bibfnamefont{D.~C.} \bibnamefont{Ralph}}, \bibnamefont{and}
  \bibinfo{author}{\bibfnamefont{R.~A.} \bibnamefont{Buhrman}},
  \bibinfo{journal}{Applied Physics Letters} \textbf{\bibinfo{volume}{101}},
  \bibinfo{pages}{122404} (\bibinfo{year}{2012}),
  \eprint{https://doi.org/10.1063/1.4753947},
  \urlprefix\url{https://doi.org/10.1063/1.4753947}.

\bibitem[{\citenamefont{Balakrishnan et~al.}(2013)\citenamefont{Balakrishnan,
  Kok Wai~Koon, Jaiswal, Castro~Neto, and Ozyilmaz}}]{10.1038/nphys2576}
\bibinfo{author}{\bibfnamefont{J.}~\bibnamefont{Balakrishnan}},
  \bibinfo{author}{\bibfnamefont{G.}~\bibnamefont{Kok Wai~Koon}},
  \bibinfo{author}{\bibfnamefont{M.}~\bibnamefont{Jaiswal}},
  \bibinfo{author}{\bibfnamefont{A.~H.} \bibnamefont{Castro~Neto}},
  \bibnamefont{and} \bibinfo{author}{\bibfnamefont{B.}~\bibnamefont{Ozyilmaz}},
  \bibinfo{journal}{Nature Physics} \textbf{\bibinfo{volume}{9}},
  \bibinfo{pages}{284} (\bibinfo{year}{2013}),
  \urlprefix\url{https://doi.org/10.1038/nphys2576}.

\bibitem[{\citenamefont{Balakrishnan et~al.}(2014)\citenamefont{Balakrishnan,
  Koon, Avsar, Ho, Lee, Jaiswal, Baeck, Ahn, Ferreira, Cazalilla
  et~al.}}]{10.1038/ncomms5748}
\bibinfo{author}{\bibfnamefont{J.}~\bibnamefont{Balakrishnan}},
  \bibinfo{author}{\bibfnamefont{G.~K.~W.} \bibnamefont{Koon}},
  \bibinfo{author}{\bibfnamefont{A.}~\bibnamefont{Avsar}},
  \bibinfo{author}{\bibfnamefont{Y.}~\bibnamefont{Ho}},
  \bibinfo{author}{\bibfnamefont{J.~H.} \bibnamefont{Lee}},
  \bibinfo{author}{\bibfnamefont{M.}~\bibnamefont{Jaiswal}},
  \bibinfo{author}{\bibfnamefont{S.-J.} \bibnamefont{Baeck}},
  \bibinfo{author}{\bibfnamefont{J.-H.} \bibnamefont{Ahn}},
  \bibinfo{author}{\bibfnamefont{A.}~\bibnamefont{Ferreira}},
  \bibinfo{author}{\bibfnamefont{M.~A.} \bibnamefont{Cazalilla}},
  \bibnamefont{et~al.}, \bibinfo{journal}{Nature Communications}
  \textbf{\bibinfo{volume}{102}}, \bibinfo{pages}{4748} (\bibinfo{year}{2014}),
  \urlprefix\url{https://doi.org/10.1038/ncomms5748}.

\bibitem[{\citenamefont{{Rappoport}}(2023)}]{2023Natur.619...38R}
\bibinfo{author}{\bibfnamefont{T.~G.} \bibnamefont{{Rappoport}}},
  \bibinfo{journal}{\nat} \textbf{\bibinfo{volume}{619}}, \bibinfo{pages}{38}
  (\bibinfo{year}{2023}).

\bibitem[{\citenamefont{Bernevig et~al.}(2005)\citenamefont{Bernevig, Hughes,
  and Zhang}}]{PhysRevLett.95.066601}
\bibinfo{author}{\bibfnamefont{B.~A.} \bibnamefont{Bernevig}},
  \bibinfo{author}{\bibfnamefont{T.~L.} \bibnamefont{Hughes}},
  \bibnamefont{and} \bibinfo{author}{\bibfnamefont{S.-C.} \bibnamefont{Zhang}},
  \bibinfo{journal}{Phys. Rev. Lett.} \textbf{\bibinfo{volume}{95}},
  \bibinfo{pages}{066601} (\bibinfo{year}{2005}),
  \urlprefix\url{https://link.aps.org/doi/10.1103/PhysRevLett.95.066601}.

\bibitem[{\citenamefont{Tanaka et~al.}(2008)\citenamefont{Tanaka, Kontani,
  Naito, Naito, Hirashima, Yamada, and Inoue}}]{PhysRevB.77.165117}
\bibinfo{author}{\bibfnamefont{T.}~\bibnamefont{Tanaka}},
  \bibinfo{author}{\bibfnamefont{H.}~\bibnamefont{Kontani}},
  \bibinfo{author}{\bibfnamefont{M.}~\bibnamefont{Naito}},
  \bibinfo{author}{\bibfnamefont{T.}~\bibnamefont{Naito}},
  \bibinfo{author}{\bibfnamefont{D.~S.} \bibnamefont{Hirashima}},
  \bibinfo{author}{\bibfnamefont{K.}~\bibnamefont{Yamada}}, \bibnamefont{and}
  \bibinfo{author}{\bibfnamefont{J.}~\bibnamefont{Inoue}},
  \bibinfo{journal}{Phys. Rev. B} \textbf{\bibinfo{volume}{77}},
  \bibinfo{pages}{165117} (\bibinfo{year}{2008}),
  \urlprefix\url{https://link.aps.org/doi/10.1103/PhysRevB.77.165117}.

\bibitem[{\citenamefont{Phong et~al.}(2019)\citenamefont{Phong, Addison, Ahn,
  Min, Agarwal, and Mele}}]{PhysRevLett.123.236403}
\bibinfo{author}{\bibfnamefont{V.~o.~T.} \bibnamefont{Phong}},
  \bibinfo{author}{\bibfnamefont{Z.}~\bibnamefont{Addison}},
  \bibinfo{author}{\bibfnamefont{S.}~\bibnamefont{Ahn}},
  \bibinfo{author}{\bibfnamefont{H.}~\bibnamefont{Min}},
  \bibinfo{author}{\bibfnamefont{R.}~\bibnamefont{Agarwal}}, \bibnamefont{and}
  \bibinfo{author}{\bibfnamefont{E.~J.} \bibnamefont{Mele}},
  \bibinfo{journal}{Phys. Rev. Lett.} \textbf{\bibinfo{volume}{123}},
  \bibinfo{pages}{236403} (\bibinfo{year}{2019}),
  \urlprefix\url{https://link.aps.org/doi/10.1103/PhysRevLett.123.236403}.

\bibitem[{\citenamefont{Go et~al.}(2018)\citenamefont{Go, Jo, Kim, and
  Lee}}]{PhysRevLett.121.086602}
\bibinfo{author}{\bibfnamefont{D.}~\bibnamefont{Go}},
  \bibinfo{author}{\bibfnamefont{D.}~\bibnamefont{Jo}},
  \bibinfo{author}{\bibfnamefont{C.}~\bibnamefont{Kim}}, \bibnamefont{and}
  \bibinfo{author}{\bibfnamefont{H.-W.} \bibnamefont{Lee}},
  \bibinfo{journal}{Phys. Rev. Lett.} \textbf{\bibinfo{volume}{121}},
  \bibinfo{pages}{086602} (\bibinfo{year}{2018}),
  \urlprefix\url{https://link.aps.org/doi/10.1103/PhysRevLett.121.086602}.

\bibitem[{\citenamefont{Jo et~al.}(2018)\citenamefont{Jo, Go, and
  Lee}}]{PhysRevB.98.214405}
\bibinfo{author}{\bibfnamefont{D.}~\bibnamefont{Jo}},
  \bibinfo{author}{\bibfnamefont{D.}~\bibnamefont{Go}}, \bibnamefont{and}
  \bibinfo{author}{\bibfnamefont{H.-W.} \bibnamefont{Lee}},
  \bibinfo{journal}{Phys. Rev. B} \textbf{\bibinfo{volume}{98}},
  \bibinfo{pages}{214405} (\bibinfo{year}{2018}),
  \urlprefix\url{https://link.aps.org/doi/10.1103/PhysRevB.98.214405}.

\bibitem[{\citenamefont{Salemi et~al.}(2021)\citenamefont{Salemi, Berritta, and
  Oppeneer}}]{PhysRevMaterials.5.074407}
\bibinfo{author}{\bibfnamefont{L.}~\bibnamefont{Salemi}},
  \bibinfo{author}{\bibfnamefont{M.}~\bibnamefont{Berritta}}, \bibnamefont{and}
  \bibinfo{author}{\bibfnamefont{P.~M.} \bibnamefont{Oppeneer}},
  \bibinfo{journal}{Phys. Rev. Materials} \textbf{\bibinfo{volume}{5}},
  \bibinfo{pages}{074407} (\bibinfo{year}{2021}),
  \urlprefix\url{https://link.aps.org/doi/10.1103/PhysRevMaterials.5.074407}.

\bibitem[{\citenamefont{Canonico
  et~al.}(2020{\natexlab{a}})\citenamefont{Canonico, Cysne, Molina-Sanchez,
  Muniz, and Rappoport}}]{PhysRevB.101.161409}
\bibinfo{author}{\bibfnamefont{L.~M.} \bibnamefont{Canonico}},
  \bibinfo{author}{\bibfnamefont{T.~P.} \bibnamefont{Cysne}},
  \bibinfo{author}{\bibfnamefont{A.}~\bibnamefont{Molina-Sanchez}},
  \bibinfo{author}{\bibfnamefont{R.~B.} \bibnamefont{Muniz}}, \bibnamefont{and}
  \bibinfo{author}{\bibfnamefont{T.~G.} \bibnamefont{Rappoport}},
  \bibinfo{journal}{Phys. Rev. B} \textbf{\bibinfo{volume}{101}},
  \bibinfo{pages}{161409} (\bibinfo{year}{2020}{\natexlab{a}}),
  \urlprefix\url{https://link.aps.org/doi/10.1103/PhysRevB.101.161409}.

\bibitem[{\citenamefont{Canonico
  et~al.}(2020{\natexlab{b}})\citenamefont{Canonico, Cysne, Rappoport, and
  Muniz}}]{PhysRevB.101.075429}
\bibinfo{author}{\bibfnamefont{L.~M.} \bibnamefont{Canonico}},
  \bibinfo{author}{\bibfnamefont{T.~P.} \bibnamefont{Cysne}},
  \bibinfo{author}{\bibfnamefont{T.~G.} \bibnamefont{Rappoport}},
  \bibnamefont{and} \bibinfo{author}{\bibfnamefont{R.~B.} \bibnamefont{Muniz}},
  \bibinfo{journal}{Phys. Rev. B} \textbf{\bibinfo{volume}{101}},
  \bibinfo{pages}{075429} (\bibinfo{year}{2020}{\natexlab{b}}),
  \urlprefix\url{https://link.aps.org/doi/10.1103/PhysRevB.101.075429}.

\bibitem[{\citenamefont{Bhowal and Satpathy}(2020)}]{PhysRevB.102.035409}
\bibinfo{author}{\bibfnamefont{S.}~\bibnamefont{Bhowal}} \bibnamefont{and}
  \bibinfo{author}{\bibfnamefont{S.}~\bibnamefont{Satpathy}},
  \bibinfo{journal}{Phys. Rev. B} \textbf{\bibinfo{volume}{102}},
  \bibinfo{pages}{035409} (\bibinfo{year}{2020}),
  \urlprefix\url{https://link.aps.org/doi/10.1103/PhysRevB.102.035409}.

\bibitem[{\citenamefont{Cysne et~al.}(2021)\citenamefont{Cysne, Costa,
  Canonico, Nardelli, Muniz, and Rappoport}}]{PhysRevLett.126.056601}
\bibinfo{author}{\bibfnamefont{T.~P.} \bibnamefont{Cysne}},
  \bibinfo{author}{\bibfnamefont{M.}~\bibnamefont{Costa}},
  \bibinfo{author}{\bibfnamefont{L.~M.} \bibnamefont{Canonico}},
  \bibinfo{author}{\bibfnamefont{M.~B.} \bibnamefont{Nardelli}},
  \bibinfo{author}{\bibfnamefont{R.~B.} \bibnamefont{Muniz}}, \bibnamefont{and}
  \bibinfo{author}{\bibfnamefont{T.~G.} \bibnamefont{Rappoport}},
  \bibinfo{journal}{Phys. Rev. Lett.} \textbf{\bibinfo{volume}{126}},
  \bibinfo{pages}{056601} (\bibinfo{year}{2021}),
  \urlprefix\url{https://link.aps.org/doi/10.1103/PhysRevLett.126.056601}.

\bibitem[{\citenamefont{Costa et~al.}(2023)\citenamefont{Costa, Focassio,
  Canonico, Cysne, Schleder, Muniz, Fazzio, and
  Rappoport}}]{PhysRevLett.130.116204}
\bibinfo{author}{\bibfnamefont{M.}~\bibnamefont{Costa}},
  \bibinfo{author}{\bibfnamefont{B.}~\bibnamefont{Focassio}},
  \bibinfo{author}{\bibfnamefont{L.~M.} \bibnamefont{Canonico}},
  \bibinfo{author}{\bibfnamefont{T.~P.} \bibnamefont{Cysne}},
  \bibinfo{author}{\bibfnamefont{G.~R.} \bibnamefont{Schleder}},
  \bibinfo{author}{\bibfnamefont{R.~B.} \bibnamefont{Muniz}},
  \bibinfo{author}{\bibfnamefont{A.}~\bibnamefont{Fazzio}}, \bibnamefont{and}
  \bibinfo{author}{\bibfnamefont{T.~G.} \bibnamefont{Rappoport}},
  \bibinfo{journal}{Phys. Rev. Lett.} \textbf{\bibinfo{volume}{130}},
  \bibinfo{pages}{116204} (\bibinfo{year}{2023}),
  \urlprefix\url{https://link.aps.org/doi/10.1103/PhysRevLett.130.116204}.

\bibitem[{\citenamefont{Sahu et~al.}(2021)\citenamefont{Sahu, Bhowal, and
  Satpathy}}]{PhysRevB.103.085113}
\bibinfo{author}{\bibfnamefont{P.}~\bibnamefont{Sahu}},
  \bibinfo{author}{\bibfnamefont{S.}~\bibnamefont{Bhowal}}, \bibnamefont{and}
  \bibinfo{author}{\bibfnamefont{S.}~\bibnamefont{Satpathy}},
  \bibinfo{journal}{Phys. Rev. B} \textbf{\bibinfo{volume}{103}},
  \bibinfo{pages}{085113} (\bibinfo{year}{2021}),
  \urlprefix\url{https://link.aps.org/doi/10.1103/PhysRevB.103.085113}.

\bibitem[{\citenamefont{Bhowal and Vignale}(2021)}]{PhysRevB.103.195309}
\bibinfo{author}{\bibfnamefont{S.}~\bibnamefont{Bhowal}} \bibnamefont{and}
  \bibinfo{author}{\bibfnamefont{G.}~\bibnamefont{Vignale}},
  \bibinfo{journal}{Phys. Rev. B} \textbf{\bibinfo{volume}{103}},
  \bibinfo{pages}{195309} (\bibinfo{year}{2021}),
  \urlprefix\url{https://link.aps.org/doi/10.1103/PhysRevB.103.195309}.

\bibitem[{\citenamefont{Go and Lee}(2020)}]{PhysRevResearch.2.013177}
\bibinfo{author}{\bibfnamefont{D.}~\bibnamefont{Go}} \bibnamefont{and}
  \bibinfo{author}{\bibfnamefont{H.-W.} \bibnamefont{Lee}},
  \bibinfo{journal}{Phys. Rev. Research} \textbf{\bibinfo{volume}{2}},
  \bibinfo{pages}{013177} (\bibinfo{year}{2020}),
  \urlprefix\url{https://link.aps.org/doi/10.1103/PhysRevResearch.2.013177}.

\bibitem[{\citenamefont{Ding et~al.}(2022)\citenamefont{Ding, Liang, Go, Yun,
  Xue, Liu, Becker, Yang, Du, Wang et~al.}}]{PhysRevLett.128.067201}
\bibinfo{author}{\bibfnamefont{S.}~\bibnamefont{Ding}},
  \bibinfo{author}{\bibfnamefont{Z.}~\bibnamefont{Liang}},
  \bibinfo{author}{\bibfnamefont{D.}~\bibnamefont{Go}},
  \bibinfo{author}{\bibfnamefont{C.}~\bibnamefont{Yun}},
  \bibinfo{author}{\bibfnamefont{M.}~\bibnamefont{Xue}},
  \bibinfo{author}{\bibfnamefont{Z.}~\bibnamefont{Liu}},
  \bibinfo{author}{\bibfnamefont{S.}~\bibnamefont{Becker}},
  \bibinfo{author}{\bibfnamefont{W.}~\bibnamefont{Yang}},
  \bibinfo{author}{\bibfnamefont{H.}~\bibnamefont{Du}},
  \bibinfo{author}{\bibfnamefont{C.}~\bibnamefont{Wang}}, \bibnamefont{et~al.},
  \bibinfo{journal}{Phys. Rev. Lett.} \textbf{\bibinfo{volume}{128}},
  \bibinfo{pages}{067201} (\bibinfo{year}{2022}),
  \urlprefix\url{https://link.aps.org/doi/10.1103/PhysRevLett.128.067201}.

\bibitem[{\citenamefont{Liao et~al.}(2022)\citenamefont{Liao, Xue, Han, Kim,
  Zhang, Li, Liu, Kou, Song, Pan et~al.}}]{PhysRevB.105.104434}
\bibinfo{author}{\bibfnamefont{L.}~\bibnamefont{Liao}},
  \bibinfo{author}{\bibfnamefont{F.}~\bibnamefont{Xue}},
  \bibinfo{author}{\bibfnamefont{L.}~\bibnamefont{Han}},
  \bibinfo{author}{\bibfnamefont{J.}~\bibnamefont{Kim}},
  \bibinfo{author}{\bibfnamefont{R.}~\bibnamefont{Zhang}},
  \bibinfo{author}{\bibfnamefont{L.}~\bibnamefont{Li}},
  \bibinfo{author}{\bibfnamefont{J.}~\bibnamefont{Liu}},
  \bibinfo{author}{\bibfnamefont{X.}~\bibnamefont{Kou}},
  \bibinfo{author}{\bibfnamefont{C.}~\bibnamefont{Song}},
  \bibinfo{author}{\bibfnamefont{F.}~\bibnamefont{Pan}}, \bibnamefont{et~al.},
  \bibinfo{journal}{Phys. Rev. B} \textbf{\bibinfo{volume}{105}},
  \bibinfo{pages}{104434} (\bibinfo{year}{2022}),
  \urlprefix\url{https://link.aps.org/doi/10.1103/PhysRevB.105.104434}.

\bibitem[{\citenamefont{Ding et~al.}(2020)\citenamefont{Ding, Ross, Go,
  Baldrati, Ren, Freimuth, Becker, Kammerbauer, Yang, Jakob
  et~al.}}]{PhysRevLett.125.177201}
\bibinfo{author}{\bibfnamefont{S.}~\bibnamefont{Ding}},
  \bibinfo{author}{\bibfnamefont{A.}~\bibnamefont{Ross}},
  \bibinfo{author}{\bibfnamefont{D.}~\bibnamefont{Go}},
  \bibinfo{author}{\bibfnamefont{L.}~\bibnamefont{Baldrati}},
  \bibinfo{author}{\bibfnamefont{Z.}~\bibnamefont{Ren}},
  \bibinfo{author}{\bibfnamefont{F.}~\bibnamefont{Freimuth}},
  \bibinfo{author}{\bibfnamefont{S.}~\bibnamefont{Becker}},
  \bibinfo{author}{\bibfnamefont{F.}~\bibnamefont{Kammerbauer}},
  \bibinfo{author}{\bibfnamefont{J.}~\bibnamefont{Yang}},
  \bibinfo{author}{\bibfnamefont{G.}~\bibnamefont{Jakob}},
  \bibnamefont{et~al.}, \bibinfo{journal}{Phys. Rev. Lett.}
  \textbf{\bibinfo{volume}{125}}, \bibinfo{pages}{177201}
  (\bibinfo{year}{2020}),
  \urlprefix\url{https://link.aps.org/doi/10.1103/PhysRevLett.125.177201}.

\bibitem[{\citenamefont{Salvador-Sánchez
  et~al.}(2022)\citenamefont{Salvador-Sánchez, Canonico, Pérez-Rodríguez,
  Cysne, Baba, Clericò, Vila, Vaquero, Delgado-Notario, Caridad
  et~al.}}]{salvadorsanchez2022generation}
\bibinfo{author}{\bibfnamefont{J.}~\bibnamefont{Salvador-Sánchez}},
  \bibinfo{author}{\bibfnamefont{L.~M.} \bibnamefont{Canonico}},
  \bibinfo{author}{\bibfnamefont{A.}~\bibnamefont{Pérez-Rodríguez}},
  \bibinfo{author}{\bibfnamefont{T.~P.} \bibnamefont{Cysne}},
  \bibinfo{author}{\bibfnamefont{Y.}~\bibnamefont{Baba}},
  \bibinfo{author}{\bibfnamefont{V.}~\bibnamefont{Clericò}},
  \bibinfo{author}{\bibfnamefont{M.}~\bibnamefont{Vila}},
  \bibinfo{author}{\bibfnamefont{D.}~\bibnamefont{Vaquero}},
  \bibinfo{author}{\bibfnamefont{J.~A.} \bibnamefont{Delgado-Notario}},
  \bibinfo{author}{\bibfnamefont{J.~M.} \bibnamefont{Caridad}},
  \bibnamefont{et~al.}, \emph{\bibinfo{title}{Generation and control of
  non-local chiral currents in graphene superlattices by orbital hall effect}}
  (\bibinfo{year}{2022}), \eprint{2206.04565},
  \urlprefix\url{https://doi.org/10.48550/arXiv.2206.04565}.

\bibitem[{\citenamefont{Han et~al.}(2022)\citenamefont{Han, Lee, and
  Kim}}]{PhysRevLett.128.176601}
\bibinfo{author}{\bibfnamefont{S.}~\bibnamefont{Han}},
  \bibinfo{author}{\bibfnamefont{H.-W.} \bibnamefont{Lee}}, \bibnamefont{and}
  \bibinfo{author}{\bibfnamefont{K.-W.} \bibnamefont{Kim}},
  \bibinfo{journal}{Phys. Rev. Lett.} \textbf{\bibinfo{volume}{128}},
  \bibinfo{pages}{176601} (\bibinfo{year}{2022}),
  \urlprefix\url{https://link.aps.org/doi/10.1103/PhysRevLett.128.176601}.

\bibitem[{\citenamefont{Lee et~al.}(2021)\citenamefont{Lee, Kang, Go, Kim,
  Kang, Lee, Lee, Kang, Lee, Mokrousov et~al.}}]{Lee2021}
\bibinfo{author}{\bibfnamefont{S.}~\bibnamefont{Lee}},
  \bibinfo{author}{\bibfnamefont{M.-G.} \bibnamefont{Kang}},
  \bibinfo{author}{\bibfnamefont{D.}~\bibnamefont{Go}},
  \bibinfo{author}{\bibfnamefont{D.}~\bibnamefont{Kim}},
  \bibinfo{author}{\bibfnamefont{J.-H.} \bibnamefont{Kang}},
  \bibinfo{author}{\bibfnamefont{T.}~\bibnamefont{Lee}},
  \bibinfo{author}{\bibfnamefont{G.-H.} \bibnamefont{Lee}},
  \bibinfo{author}{\bibfnamefont{J.}~\bibnamefont{Kang}},
  \bibinfo{author}{\bibfnamefont{N.~J.} \bibnamefont{Lee}},
  \bibinfo{author}{\bibfnamefont{Y.}~\bibnamefont{Mokrousov}},
  \bibnamefont{et~al.}, \bibinfo{journal}{Communications Physics}
  \textbf{\bibinfo{volume}{4}}, \bibinfo{pages}{234} (\bibinfo{year}{2021}),
  \urlprefix\url{https://doi.org/10.1038/s42005-021-00737-7}.

\bibitem[{\citenamefont{Choi et~al.}(2023)\citenamefont{Choi, Jo, Ko, Go, Kim,
  Park, Kim, Min, Choi, and Lee}}]{https://doi.org/10.48550/arxiv.2109.14847}
\bibinfo{author}{\bibfnamefont{Y.-G.} \bibnamefont{Choi}},
  \bibinfo{author}{\bibfnamefont{D.}~\bibnamefont{Jo}},
  \bibinfo{author}{\bibfnamefont{K.-H.} \bibnamefont{Ko}},
  \bibinfo{author}{\bibfnamefont{D.}~\bibnamefont{Go}},
  \bibinfo{author}{\bibfnamefont{K.-H.} \bibnamefont{Kim}},
  \bibinfo{author}{\bibfnamefont{H.~G.} \bibnamefont{Park}},
  \bibinfo{author}{\bibfnamefont{C.}~\bibnamefont{Kim}},
  \bibinfo{author}{\bibfnamefont{B.-C.} \bibnamefont{Min}},
  \bibinfo{author}{\bibfnamefont{G.-M.} \bibnamefont{Choi}}, \bibnamefont{and}
  \bibinfo{author}{\bibfnamefont{H.-W.} \bibnamefont{Lee}},
  \bibinfo{journal}{Nature} \textbf{\bibinfo{volume}{619}}, \bibinfo{pages}{52}
  (\bibinfo{year}{2023}),
  \urlprefix\url{https://doi.org/10.1038%2Fs41586-023-06101-9}.

\bibitem[{\citenamefont{Hayashi et~al.}(2023)\citenamefont{Hayashi, Jo, Go,
  Gao, Haku, Mokrousov, Lee, and
  Ando}}]{https://doi.org/10.48550/arxiv.2202.13896}
\bibinfo{author}{\bibfnamefont{H.}~\bibnamefont{Hayashi}},
  \bibinfo{author}{\bibfnamefont{D.}~\bibnamefont{Jo}},
  \bibinfo{author}{\bibfnamefont{D.}~\bibnamefont{Go}},
  \bibinfo{author}{\bibfnamefont{T.}~\bibnamefont{Gao}},
  \bibinfo{author}{\bibfnamefont{S.}~\bibnamefont{Haku}},
  \bibinfo{author}{\bibfnamefont{Y.}~\bibnamefont{Mokrousov}},
  \bibinfo{author}{\bibfnamefont{H.-W.} \bibnamefont{Lee}}, \bibnamefont{and}
  \bibinfo{author}{\bibfnamefont{K.}~\bibnamefont{Ando}},
  \bibinfo{journal}{Communications Physics} \textbf{\bibinfo{volume}{6}}
  (\bibinfo{year}{2023}),
  \urlprefix\url{https://doi.org/10.1038%2Fs42005-023-01139-7}.

\bibitem[{\citenamefont{Sala and Gambardella}(2022)}]{PhysRevResearch.4.033037}
\bibinfo{author}{\bibfnamefont{G.}~\bibnamefont{Sala}} \bibnamefont{and}
  \bibinfo{author}{\bibfnamefont{P.}~\bibnamefont{Gambardella}},
  \bibinfo{journal}{Phys. Rev. Res.} \textbf{\bibinfo{volume}{4}},
  \bibinfo{pages}{033037} (\bibinfo{year}{2022}),
  \urlprefix\url{https://link.aps.org/doi/10.1103/PhysRevResearch.4.033037}.

\bibitem[{\citenamefont{Santos et~al.}(2023)\citenamefont{Santos, Abr\~ao, Go,
  de~Assis, Mokrousov, Mendes, and
  Azevedo}}]{https://doi.org/10.48550/arxiv.2204.01825}
\bibinfo{author}{\bibfnamefont{E.}~\bibnamefont{Santos}},
  \bibinfo{author}{\bibfnamefont{J.}~\bibnamefont{Abr\~ao}},
  \bibinfo{author}{\bibfnamefont{D.}~\bibnamefont{Go}},
  \bibinfo{author}{\bibfnamefont{L.}~\bibnamefont{de~Assis}},
  \bibinfo{author}{\bibfnamefont{Y.}~\bibnamefont{Mokrousov}},
  \bibinfo{author}{\bibfnamefont{J.}~\bibnamefont{Mendes}}, \bibnamefont{and}
  \bibinfo{author}{\bibfnamefont{A.}~\bibnamefont{Azevedo}},
  \bibinfo{journal}{Phys. Rev. Appl.} \textbf{\bibinfo{volume}{19}},
  \bibinfo{pages}{014069} (\bibinfo{year}{2023}),
  \urlprefix\url{https://link.aps.org/doi/10.1103/PhysRevApplied.19.014069}.

\bibitem[{\citenamefont{Baek and Lee}(2021)}]{PhysRevB.104.245204}
\bibinfo{author}{\bibfnamefont{I.}~\bibnamefont{Baek}} \bibnamefont{and}
  \bibinfo{author}{\bibfnamefont{H.-W.} \bibnamefont{Lee}},
  \bibinfo{journal}{Phys. Rev. B} \textbf{\bibinfo{volume}{104}},
  \bibinfo{pages}{245204} (\bibinfo{year}{2021}),
  \urlprefix\url{https://link.aps.org/doi/10.1103/PhysRevB.104.245204}.

\bibitem[{\citenamefont{Bose et~al.}(2023)\citenamefont{Bose, Kammerbauer,
  Gupta, Go, Mokrousov, Jakob, and Kl\"aui}}]{PhysRevB.107.134423}
\bibinfo{author}{\bibfnamefont{A.}~\bibnamefont{Bose}},
  \bibinfo{author}{\bibfnamefont{F.}~\bibnamefont{Kammerbauer}},
  \bibinfo{author}{\bibfnamefont{R.}~\bibnamefont{Gupta}},
  \bibinfo{author}{\bibfnamefont{D.}~\bibnamefont{Go}},
  \bibinfo{author}{\bibfnamefont{Y.}~\bibnamefont{Mokrousov}},
  \bibinfo{author}{\bibfnamefont{G.}~\bibnamefont{Jakob}}, \bibnamefont{and}
  \bibinfo{author}{\bibfnamefont{M.}~\bibnamefont{Kl\"aui}},
  \bibinfo{journal}{Phys. Rev. B} \textbf{\bibinfo{volume}{107}},
  \bibinfo{pages}{134423} (\bibinfo{year}{2023}),
  \urlprefix\url{https://link.aps.org/doi/10.1103/PhysRevB.107.134423}.

\bibitem[{\citenamefont{Go et~al.}(2021)\citenamefont{Go, Jo, Lee, Kläui, and
  Mokrousov}}]{Go_2021}
\bibinfo{author}{\bibfnamefont{D.}~\bibnamefont{Go}},
  \bibinfo{author}{\bibfnamefont{D.}~\bibnamefont{Jo}},
  \bibinfo{author}{\bibfnamefont{H.-W.} \bibnamefont{Lee}},
  \bibinfo{author}{\bibfnamefont{M.}~\bibnamefont{Kläui}}, \bibnamefont{and}
  \bibinfo{author}{\bibfnamefont{Y.}~\bibnamefont{Mokrousov}},
  \bibinfo{journal}{Europhysics Letters} \textbf{\bibinfo{volume}{135}},
  \bibinfo{pages}{37001} (\bibinfo{year}{2021}),
  \urlprefix\url{https://dx.doi.org/10.1209/0295-5075/ac2653}.

\bibitem[{\citenamefont{Kim and Otani}(2022)}]{KIM2022169974}
\bibinfo{author}{\bibfnamefont{J.}~\bibnamefont{Kim}} \bibnamefont{and}
  \bibinfo{author}{\bibfnamefont{Y.}~\bibnamefont{Otani}},
  \bibinfo{journal}{Journal of Magnetism and Magnetic Materials}
  \textbf{\bibinfo{volume}{563}}, \bibinfo{pages}{169974}
  (\bibinfo{year}{2022}), ISSN \bibinfo{issn}{0304-8853},
  \urlprefix\url{https://www.sciencedirect.com/science/article/pii/S0304885322008599}.

\bibitem[{\citenamefont{Washburn and
  Webb}(1986)}]{doi:10.1080/00018738600101921}
\bibinfo{author}{\bibfnamefont{S.}~\bibnamefont{Washburn}} \bibnamefont{and}
  \bibinfo{author}{\bibfnamefont{R.~A.} \bibnamefont{Webb}},
  \bibinfo{journal}{Advances in Physics} \textbf{\bibinfo{volume}{35}},
  \bibinfo{pages}{375} (\bibinfo{year}{1986}),
  \eprint{https://doi.org/10.1080/00018738600101921},
  \urlprefix\url{https://doi.org/10.1080/00018738600101921}.

\bibitem[{\citenamefont{Beenakker}(1997)}]{RevModPhys.69.731}
\bibinfo{author}{\bibfnamefont{C.~W.~J.} \bibnamefont{Beenakker}},
  \bibinfo{journal}{Rev. Mod. Phys.} \textbf{\bibinfo{volume}{69}},
  \bibinfo{pages}{731} (\bibinfo{year}{1997}),
  \urlprefix\url{https://link.aps.org/doi/10.1103/RevModPhys.69.731}.

\bibitem[{\citenamefont{Ren et~al.}(2006)\citenamefont{Ren, Qiao, Wang, Sun,
  and Guo}}]{PhysRevLett.97.066603}
\bibinfo{author}{\bibfnamefont{W.}~\bibnamefont{Ren}},
  \bibinfo{author}{\bibfnamefont{Z.}~\bibnamefont{Qiao}},
  \bibinfo{author}{\bibfnamefont{J.}~\bibnamefont{Wang}},
  \bibinfo{author}{\bibfnamefont{Q.}~\bibnamefont{Sun}}, \bibnamefont{and}
  \bibinfo{author}{\bibfnamefont{H.}~\bibnamefont{Guo}},
  \bibinfo{journal}{Phys. Rev. Lett.} \textbf{\bibinfo{volume}{97}},
  \bibinfo{pages}{066603} (\bibinfo{year}{2006}),
  \urlprefix\url{https://link.aps.org/doi/10.1103/PhysRevLett.97.066603}.

\bibitem[{\citenamefont{Bardarson et~al.}(2007)\citenamefont{Bardarson,
  Adagideli, and Jacquod}}]{PhysRevLett.98.196601}
\bibinfo{author}{\bibfnamefont{J.~H.} \bibnamefont{Bardarson}},
  \bibinfo{author}{\bibfnamefont{i.~d.~I.} \bibnamefont{Adagideli}},
  \bibnamefont{and} \bibinfo{author}{\bibfnamefont{P.}~\bibnamefont{Jacquod}},
  \bibinfo{journal}{Phys. Rev. Lett.} \textbf{\bibinfo{volume}{98}},
  \bibinfo{pages}{196601} (\bibinfo{year}{2007}),
  \urlprefix\url{https://link.aps.org/doi/10.1103/PhysRevLett.98.196601}.

\bibitem[{\citenamefont{Saitoh et~al.}(2006)\citenamefont{Saitoh, Ueda,
  Miyajima, and Tatara}}]{doi:10.1063/1.2199473}
\bibinfo{author}{\bibfnamefont{E.}~\bibnamefont{Saitoh}},
  \bibinfo{author}{\bibfnamefont{M.}~\bibnamefont{Ueda}},
  \bibinfo{author}{\bibfnamefont{H.}~\bibnamefont{Miyajima}}, \bibnamefont{and}
  \bibinfo{author}{\bibfnamefont{G.}~\bibnamefont{Tatara}},
  \bibinfo{journal}{Applied Physics Letters} \textbf{\bibinfo{volume}{88}},
  \bibinfo{pages}{182509} (\bibinfo{year}{2006}),
  \eprint{https://doi.org/10.1063/1.2199473},
  \urlprefix\url{https://doi.org/10.1063/1.2199473}.

\bibitem[{\citenamefont{Azevedo et~al.}(2005)\citenamefont{Azevedo,
  Vilela~Leão, Rodriguez-Suarez, Oliveira, and
  Rezende}}]{doi:10.1063/1.1855251}
\bibinfo{author}{\bibfnamefont{A.}~\bibnamefont{Azevedo}},
  \bibinfo{author}{\bibfnamefont{L.~H.} \bibnamefont{Vilela~Leão}},
  \bibinfo{author}{\bibfnamefont{R.~L.} \bibnamefont{Rodriguez-Suarez}},
  \bibinfo{author}{\bibfnamefont{A.~B.} \bibnamefont{Oliveira}},
  \bibnamefont{and} \bibinfo{author}{\bibfnamefont{S.~M.}
  \bibnamefont{Rezende}}, \bibinfo{journal}{Journal of Applied Physics}
  \textbf{\bibinfo{volume}{97}}, \bibinfo{pages}{10C715}
  (\bibinfo{year}{2005}), \eprint{https://doi.org/10.1063/1.1855251},
  \urlprefix\url{https://doi.org/10.1063/1.1855251}.

\bibitem[{\citenamefont{Mendes et~al.}(2015)\citenamefont{Mendes, Alves~Santos,
  Meireles, Lacerda, Vilela-Le\~ao, Machado, Rodr\'{\i}guez-Su\'arez, Azevedo,
  and Rezende}}]{PhysRevLett.115.226601}
\bibinfo{author}{\bibfnamefont{J.~B.~S.} \bibnamefont{Mendes}},
  \bibinfo{author}{\bibfnamefont{O.}~\bibnamefont{Alves~Santos}},
  \bibinfo{author}{\bibfnamefont{L.~M.} \bibnamefont{Meireles}},
  \bibinfo{author}{\bibfnamefont{R.~G.} \bibnamefont{Lacerda}},
  \bibinfo{author}{\bibfnamefont{L.~H.} \bibnamefont{Vilela-Le\~ao}},
  \bibinfo{author}{\bibfnamefont{F.~L.~A.} \bibnamefont{Machado}},
  \bibinfo{author}{\bibfnamefont{R.~L.} \bibnamefont{Rodr\'{\i}guez-Su\'arez}},
  \bibinfo{author}{\bibfnamefont{A.}~\bibnamefont{Azevedo}}, \bibnamefont{and}
  \bibinfo{author}{\bibfnamefont{S.~M.} \bibnamefont{Rezende}},
  \bibinfo{journal}{Phys. Rev. Lett.} \textbf{\bibinfo{volume}{115}},
  \bibinfo{pages}{226601} (\bibinfo{year}{2015}),
  \urlprefix\url{https://link.aps.org/doi/10.1103/PhysRevLett.115.226601}.

\bibitem[{\citenamefont{Ramos et~al.}(2018)\citenamefont{Ramos, Vasconcelos,
  and Barbosa}}]{doi:10.1063/1.5010973}
\bibinfo{author}{\bibfnamefont{J.~G. G.~S.} \bibnamefont{Ramos}},
  \bibinfo{author}{\bibfnamefont{T.~C.} \bibnamefont{Vasconcelos}},
  \bibnamefont{and} \bibinfo{author}{\bibfnamefont{A.~L.~R.}
  \bibnamefont{Barbosa}}, \bibinfo{journal}{Journal of Applied Physics}
  \textbf{\bibinfo{volume}{123}}, \bibinfo{pages}{034304}
  (\bibinfo{year}{2018}), \eprint{https://doi.org/10.1063/1.5010973},
  \urlprefix\url{https://doi.org/10.1063/1.5010973}.

\bibitem[{\citenamefont{Santana et~al.}(2020)\citenamefont{Santana, da~Silva,
  Vasconcelos, Ramos, and Barbosa}}]{PhysRevB.102.041107}
\bibinfo{author}{\bibfnamefont{F.~A.~F.} \bibnamefont{Santana}},
  \bibinfo{author}{\bibfnamefont{J.~M.} \bibnamefont{da~Silva}},
  \bibinfo{author}{\bibfnamefont{T.~C.} \bibnamefont{Vasconcelos}},
  \bibinfo{author}{\bibfnamefont{J.~G. G.~S.} \bibnamefont{Ramos}},
  \bibnamefont{and} \bibinfo{author}{\bibfnamefont{A.~L.~R.}
  \bibnamefont{Barbosa}}, \bibinfo{journal}{Phys. Rev. B}
  \textbf{\bibinfo{volume}{102}}, \bibinfo{pages}{041107}
  (\bibinfo{year}{2020}),
  \urlprefix\url{https://link.aps.org/doi/10.1103/PhysRevB.102.041107}.

\bibitem[{\citenamefont{da~Silva et~al.}(2022)\citenamefont{da~Silva, Santana,
  Ramos, and Barbosa}}]{doi:10.1063/5.0107212}
\bibinfo{author}{\bibfnamefont{J.~M.} \bibnamefont{da~Silva}},
  \bibinfo{author}{\bibfnamefont{F.~A.~F.} \bibnamefont{Santana}},
  \bibinfo{author}{\bibfnamefont{J.~G. G.~S.} \bibnamefont{Ramos}},
  \bibnamefont{and} \bibinfo{author}{\bibfnamefont{A.~L.~R.}
  \bibnamefont{Barbosa}}, \bibinfo{journal}{Journal of Applied Physics}
  \textbf{\bibinfo{volume}{132}}, \bibinfo{pages}{183901}
  (\bibinfo{year}{2022}), \eprint{https://doi.org/10.1063/5.0107212},
  \urlprefix\url{https://doi.org/10.1063/5.0107212}.

\bibitem[{\citenamefont{Nikolić and Zârbo}(2007)}]{Nikolic_2007}
\bibinfo{author}{\bibfnamefont{B.~K.} \bibnamefont{Nikolić}} \bibnamefont{and}
  \bibinfo{author}{\bibfnamefont{L.~P.} \bibnamefont{Zârbo}},
  \bibinfo{journal}{Europhysics Letters} \textbf{\bibinfo{volume}{77}},
  \bibinfo{pages}{47004} (\bibinfo{year}{2007}),
  \urlprefix\url{https://dx.doi.org/10.1209/0295-5075/77/47004}.

\bibitem[{\citenamefont{Brouwer and Beenakker}(1996)}]{doi:10.1063/1.531667}
\bibinfo{author}{\bibfnamefont{P.~W.} \bibnamefont{Brouwer}} \bibnamefont{and}
  \bibinfo{author}{\bibfnamefont{C.~W.~J.} \bibnamefont{Beenakker}},
  \bibinfo{journal}{Journal of Mathematical Physics}
  \textbf{\bibinfo{volume}{37}}, \bibinfo{pages}{4904} (\bibinfo{year}{1996}),
  \eprint{https://doi.org/10.1063/1.531667},
  \urlprefix\url{https://doi.org/10.1063/1.531667}.

\bibitem[{\citenamefont{Reichl}(1980)}]{Reichl:101976}
\bibinfo{author}{\bibfnamefont{L.~E.} \bibnamefont{Reichl}},
  \emph{\bibinfo{title}{{A modern course in statistical physics}}}
  (\bibinfo{publisher}{Arnold}, \bibinfo{address}{London},
  \bibinfo{year}{1980}), \urlprefix\url{https://cds.cern.ch/record/101976}.

\bibitem[{\citenamefont{Ramos et~al.}(2012)\citenamefont{Ramos, Barbosa,
  Bazeia, Hussein, and Lewenkopf}}]{PhysRevB.86.235112}
\bibinfo{author}{\bibfnamefont{J.~G. G.~S.} \bibnamefont{Ramos}},
  \bibinfo{author}{\bibfnamefont{A.~L.~R.} \bibnamefont{Barbosa}},
  \bibinfo{author}{\bibfnamefont{D.}~\bibnamefont{Bazeia}},
  \bibinfo{author}{\bibfnamefont{M.~S.} \bibnamefont{Hussein}},
  \bibnamefont{and} \bibinfo{author}{\bibfnamefont{C.~H.}
  \bibnamefont{Lewenkopf}}, \bibinfo{journal}{Phys. Rev. B}
  \textbf{\bibinfo{volume}{86}}, \bibinfo{pages}{235112}
  (\bibinfo{year}{2012}),
  \urlprefix\url{https://link.aps.org/doi/10.1103/PhysRevB.86.235112}.

\bibitem[{\citenamefont{Vasconcelos et~al.}(2016)\citenamefont{Vasconcelos,
  Ramos, and Barbosa}}]{PhysRevB.93.115120}
\bibinfo{author}{\bibfnamefont{T.~C.} \bibnamefont{Vasconcelos}},
  \bibinfo{author}{\bibfnamefont{J.~G. G.~S.} \bibnamefont{Ramos}},
  \bibnamefont{and} \bibinfo{author}{\bibfnamefont{A.~L.~R.}
  \bibnamefont{Barbosa}}, \bibinfo{journal}{Phys. Rev. B}
  \textbf{\bibinfo{volume}{93}}, \bibinfo{pages}{115120}
  (\bibinfo{year}{2016}),
  \urlprefix\url{https://link.aps.org/doi/10.1103/PhysRevB.93.115120}.

\bibitem[{\citenamefont{Mello and Kumar}(2004)}]{Mello}
\bibinfo{author}{\bibfnamefont{P.~A.} \bibnamefont{Mello}} \bibnamefont{and}
  \bibinfo{author}{\bibfnamefont{N.}~\bibnamefont{Kumar}},
  \emph{\bibinfo{title}{{Quantum Transport in Mesoscopic Systems}}}
  (\bibinfo{publisher}{Oxford}, \bibinfo{address}{New York},
  \bibinfo{year}{2004}).

\bibitem[{\citenamefont{Mac\^edo}(2000)}]{PhysRevB.61.4453}
\bibinfo{author}{\bibfnamefont{A.~M.~S.} \bibnamefont{Mac\^edo}},
  \bibinfo{journal}{Phys. Rev. B} \textbf{\bibinfo{volume}{61}},
  \bibinfo{pages}{4453} (\bibinfo{year}{2000}),
  \urlprefix\url{https://link.aps.org/doi/10.1103/PhysRevB.61.4453}.

\bibitem[{\citenamefont{Verbaarschot et~al.}(1985)\citenamefont{Verbaarschot,
  Weidenmüller, and Zirnbauer}}]{VERBAARSCHOT1985367}
\bibinfo{author}{\bibfnamefont{J.}~\bibnamefont{Verbaarschot}},
  \bibinfo{author}{\bibfnamefont{H.}~\bibnamefont{Weidenmüller}},
  \bibnamefont{and}
  \bibinfo{author}{\bibfnamefont{M.}~\bibnamefont{Zirnbauer}},
  \bibinfo{journal}{Physics Reports} \textbf{\bibinfo{volume}{129}},
  \bibinfo{pages}{367} (\bibinfo{year}{1985}), ISSN \bibinfo{issn}{0370-1573},
  \urlprefix\url{https://www.sciencedirect.com/science/article/pii/0370157385900705}.

\bibitem[{\citenamefont{Almeida et~al.}(2009)\citenamefont{Almeida,
  Rodr\'{\i}guez-P\'erez, and Mac\^edo}}]{PhysRevB.80.125320}
\bibinfo{author}{\bibfnamefont{F.~A.~G.} \bibnamefont{Almeida}},
  \bibinfo{author}{\bibfnamefont{S.}~\bibnamefont{Rodr\'{\i}guez-P\'erez}},
  \bibnamefont{and} \bibinfo{author}{\bibfnamefont{A.~M.~S.}
  \bibnamefont{Mac\^edo}}, \bibinfo{journal}{Phys. Rev. B}
  \textbf{\bibinfo{volume}{80}}, \bibinfo{pages}{125320}
  (\bibinfo{year}{2009}),
  \urlprefix\url{https://link.aps.org/doi/10.1103/PhysRevB.80.125320}.

\bibitem[{\citenamefont{Groth et~al.}(2014)\citenamefont{Groth, Wimmer,
  Akhmerov, and Waintal}}]{kwant}
\bibinfo{author}{\bibfnamefont{C.~W.} \bibnamefont{Groth}},
  \bibinfo{author}{\bibfnamefont{M.}~\bibnamefont{Wimmer}},
  \bibinfo{author}{\bibfnamefont{A.~R.} \bibnamefont{Akhmerov}},
  \bibnamefont{and} \bibinfo{author}{\bibfnamefont{X.}~\bibnamefont{Waintal}},
  \bibinfo{journal}{New Journal of Physics} \textbf{\bibinfo{volume}{16}},
  \bibinfo{pages}{063065} (\bibinfo{year}{2014}).

\bibitem[{\citenamefont{Qiao et~al.}(2008)\citenamefont{Qiao, Wang, Wei, and
  Guo}}]{PhysRevLett.101.016804}
\bibinfo{author}{\bibfnamefont{Z.}~\bibnamefont{Qiao}},
  \bibinfo{author}{\bibfnamefont{J.}~\bibnamefont{Wang}},
  \bibinfo{author}{\bibfnamefont{Y.}~\bibnamefont{Wei}}, \bibnamefont{and}
  \bibinfo{author}{\bibfnamefont{H.}~\bibnamefont{Guo}},
  \bibinfo{journal}{Phys. Rev. Lett.} \textbf{\bibinfo{volume}{101}},
  \bibinfo{pages}{016804} (\bibinfo{year}{2008}),
  \urlprefix\url{https://link.aps.org/doi/10.1103/PhysRevLett.101.016804}.

\bibitem[{\citenamefont{Meyer et~al.}(2017)\citenamefont{Meyer, Chen, Wimmer,
  Althammer, Wimmer, Schlitz, Geprägs, Huebl, K\"odderitzsch, Ebert
  et~al.}}]{Meyer2017}
\bibinfo{author}{\bibfnamefont{S.}~\bibnamefont{Meyer}},
  \bibinfo{author}{\bibfnamefont{Y.-T.} \bibnamefont{Chen}},
  \bibinfo{author}{\bibfnamefont{S.}~\bibnamefont{Wimmer}},
  \bibinfo{author}{\bibfnamefont{M.}~\bibnamefont{Althammer}},
  \bibinfo{author}{\bibfnamefont{T.}~\bibnamefont{Wimmer}},
  \bibinfo{author}{\bibfnamefont{R.}~\bibnamefont{Schlitz}},
  \bibinfo{author}{\bibfnamefont{S.}~\bibnamefont{Geprägs}},
  \bibinfo{author}{\bibfnamefont{H.}~\bibnamefont{Huebl}},
  \bibinfo{author}{\bibfnamefont{D.}~\bibnamefont{K\"odderitzsch}},
  \bibinfo{author}{\bibfnamefont{H.}~\bibnamefont{Ebert}},
  \bibnamefont{et~al.}, \bibinfo{journal}{Nature Materials}
  \textbf{\bibinfo{volume}{16}}, \bibinfo{pages}{977} (\bibinfo{year}{2017}),
  \urlprefix\url{https://doi.org/10.1038/nmat4964}.

\end{thebibliography}

\end{document}